\documentclass{article}
\usepackage{arxiv}
\usepackage[utf8]{inputenc} 
\usepackage[T1]{fontenc}    
\usepackage{hyperref}       
\usepackage{url}            
\usepackage{booktabs}       
\usepackage{amsfonts}       
\usepackage{nicefrac}       
\usepackage{microtype}      
\usepackage{lipsum}		
\usepackage{graphicx}
\usepackage{natbib}
\usepackage{doi}

\usepackage{amsmath,amssymb,amsfonts}%

\title{Nonuniform and inequitable healthcare accessibility losses during flooding-amplified congestion across US cities}

\author{ \href{https://orcid.org/0000-0001-6689-6856}{\includegraphics[scale=0.06]{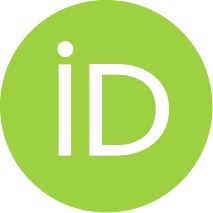}\hspace{1mm}Raviraj Dave}\thanks{Corresponding author} \\
	Sustainability and Data Sciences Laboratory, Civil and Environmental Engineering\\
	Northeastern University\\
	Boston, MA 02115 \\
	\texttt{r.dave@northeastern.edu} \\
 \And
 \href{https://orcid.org/0009-0006-7093-9672}{\includegraphics[scale=0.06]{orcid.pdf}\hspace{1mm}Danish Mansoor Tantary}\\
	Sustainability and Data Sciences Laboratory, Civil and Environmental Engineering\\
	Northeastern University\\
	Boston, MA 02115 \\
	\texttt{mansoor.d@northeastern.edu} \\
    \And
\href{https://orcid.org/0009-0008-7550-8120}{\includegraphics[scale=0.06]{orcid.pdf}\hspace{1mm}Dongqin Zhou} \\
	The Institute for Experiential AI\\
	Northeastern University\\
	Boston, MA 02115 \\
	\texttt{d.zhou@northeastern.edu} \\
 \And
 \href{https://orcid.org/0000-0002-0017-9085}{\includegraphics[scale=0.06]{orcid.pdf}\hspace{1mm}Udit Bhatia} \\
	Department of Civil Engineering, Institute of Technology Gandhinagar , Gandhinagar, Gujarat, India.\\
	Department of Earth Sciences, Institute of Technology Gandhinagar , Gandhinagar, Gujarat, India.\\
	Department of Computer Sciences and Engineering, Institute of Technology Gandhinagar , Gandhinagar, Gujarat, India. \\
	\texttt{bhatia.u@iitgn.ac.in} \\
 \And
\href{https://orcid.org/0000-0002-4292-4856}{\includegraphics[scale=0.06]{orcid.pdf}\hspace{1mm}Auroop R.~Ganguly}\thanks{Corresponding author} \\
	Sustainability and Data Sciences Laboratory, Civil and Environmental Engineering\\
    The Institute for Experiential AI\\
	Northeastern University\\
	Boston, MA 02115 \\
    Pacific Northwest National Laboratory, Richland, WA\\
	\texttt{a.ganguly@northeastern.edu} \\
}

 \renewcommand{\shorttitle}{\textit{arXiv} Template}

\begin{document}
\maketitle

\begin{abstract}
	{Cities in the United States and worldwide concentrate populations with diverse incomes, demographics and health risks, making access to healthcare facilities and emergency services critical. Urban flooding amplifies congestion and further degrades healthcare accessibility, but its impacts are uneven within and across cities, often burdening underserved populations and underplanned areas. These risks are intensifying as rapid development, aging infrastructure and climate change increase urban flood exposure. Across 30 of the most populous US cities, we show that peak-hour congestion alone reduces accessibility by 4\% to 29\%, while flooding further reduces it by 2\% to more than 90\%. These losses depend on flood type, transportation design, infrastructure planning, health vulnerability and spatial inequality. Flooding also increases spatial inequality in healthcare access, but the social burden varies by city and flood type, with urban coastal flooding disproportionately affecting less socially vulnerable populations in the United States. Our findings suggest that targeted approaches for embedding flood resilience in transportation and healthcare systems will be more effective than universal strategies, while allowing cities to learn from diverse urban regions facing similar accessibility risks.}
\end{abstract}



\section{Introduction}
Reliable access to critical facilities is fundamental to urban life, supporting community well-being and economic activity (\cite{colglazier2015sustainable,fajzel2023global,weiss2018global}). This dependence is intensifying as cities grow in population and spatial extent, placing greater pressure on the infrastructure systems that connect residents to essential services such as healthcare, education, food access, emergency response and public transport (\cite{sun2020dramatic,zeifman2022world,thacker2019infrastructure,pandey2025rising}). Among these services, healthcare is especially time-sensitive, and timely access to healthcare facilities remains essential for emergency response, routine care and public well-being (\cite{gulliford2002does,del2022modeling,baeten2018inequalities}). This access is sustained primarily by road infrastructure system in the U.S., yet the same forces that drive urban growth also increase travel demand, congestion and delay, reducing the ability of residents to reach care when needed (\cite{yiannakoulias2013estimating,levy2010evaluation,tang2018resilience,lin2024assessing,korkali2017reducing}). These concerns are particularly acute for healthcare facility access, which is highly sensitive to travel delay and often reliant on a limited set of efficient traffic corridors. Disruption of even a few critical road segments can reduce the number of reachable healthcare facilities, delay medical trips and increase pressure on already strained health systems (\cite{gangwal2022critical,schuster2024stress,papilloud2024road}).

 Climate-driven hazards add a further layer of stress by disrupting mobility when reliable access is most critical (\cite{liu2023beyond,huang2025human,garcia2025future}). Flooding is particularly consequential because it can reduce road-network capacity through direct inundation and redirect traffic across the remaining network, turning localized disruption into broader accessibility loss (\cite{dave2025local,pescaroli2018perceptions,alabbad2021assessment,serre2018assessing,rajput2023anatomy}). Globally, road and rail infrastructure are highly exposed to flooding, with expected annual damages estimated at US\$2.8–20 billion (\cite{koks2019global}). In the United States, flood risk is compounded by aging and already congested transport infrastructure (\cite{zeng2025unveiling,ASCE_IRC_2025}). About 5.36\% of the U.S. road network, or roughly 23,752 miles, is exposed to flood risk, and this exposure is expected to increase under continued warming (\cite{fathom_us_flood_risk_index_misc,koks2019global}). In urban areas where recurrent congestion already constrains movement, the interaction between flood exposure and traffic delay can sharply reshape access to healthcare, leading to missed trips, delayed evacuation services and uneven burdens across communities, as illustrated by the evacuation of Bellevue and other New York City hospitals during Hurricane Sandy in 2012 and the hospital strain, evacuation planning and patient-transfer challenges reported across Harris County during Hurricane Harvey in 2017 (\cite{yu2020disruption,wassmer2025unveiling,lin2024assessing,he2021flood,hines2023hospital,uppal2013search}). The form of this disruption also depends on the type of the flood driver. Inland cities may experience pluvial and fluvial flooding, whereas coastal cities face an additional hazard from tidal extremes (\cite{dave2025local,maymandi2022compound,wahl2015increasing,hendry2019assessing,sanders2023large}). Because these drivers differ in their spatial footprint and interaction with the road network, they can generate distinct patterns of mobility disruption and accessibility loss. Assessment based on a single flood mechanism may therefore misrepresent the resilience of transport-dependent healthcare access, especially in cities exposed to multiple interacting flood processes (\cite{sanders2023large,kasmalkar2020floods,sohrabi2025analyzing,ali2025multivariate}). The rising frequency and severity of flood disruption (\cite{frei2006future,buchanan2017amplification,slater2016recent}), together with the vital role of road transport infrastructure for societal functioning (\cite{weiss2018global,meijer2018global}) highlight the need to understand the resilience of transport-healthcare access under disrupted mobility and to inform more equitable urban infrastructure planning.

Accessibility to critical facilities has become a central concern in urban research, reflected in the growing emphasis on the concept of the “x-minute city” and related place-based approaches that evaluate how effectively residents can reach essential services within acceptable travel times (\cite{bruno2024universal,vale2023accessibility, abbiasov202415,ma202315}). Proximity-based measures and floating catchment area methods are widely used in the literature to assess accessibility by accounting for service availability, travel thresholds and, increasingly, the balance between facility supply and population demand (\cite{li2025measuring,marzolla2026proximity,chen2025public,wassmer2025unveiling,lee2025travel}). More recent advances also incorporate competition effects, recognizing that access depends not only on proximity to facilities but also on the population competing for the same services through the transport network (\cite{lin2024assessing,chen2025public}). These approaches have also been used to examine inequality in access to critical facilities across urban space (\cite{fan2022equality,verma2025spatial,michelangeli2025inequality,zhao2025unequal}). Yet many accessibility studies continue to treat travel impedance as static, overlooking the temporal variability of traffic conditions that can substantially alter realized accessibility (\cite{bruno2024universal,marzolla2026proximity,del2022modeling}). At the same time, research on flood–transport resilience has advanced through topological, functional and simulation-based approaches that capture rerouting, congestion spillback, capacity constraints and behavioural adaptation under disruption. However, the focus remains largely on network-level disruption, with much less attention to how those disruptions translate into healthcare accessibility across cities and flood conditions (\cite{wang2019local,dong2022modest,rajput2023anatomy,dave2021extreme,kasmalkar2020floods}). This is especially relevant as urban flooding can arise from distinct processes, including intense rainfall, riverine overflow, tidal surge and combination of them, each of which interacts differently with the road network (\cite{dave2025local}). Beyond this, when accessibility inequality is examined, the focus remains largely on spatial variation across zones rather than on how accessibility loss is distributed across socially vulnerable communities. Overall, these limitations constrain evaluation of how healthcare accessibility changes across the socio-econmic spectrum of a given city under temporally varying traffic and distinct flood mechanisms.

To address these challenges, we examine healthcare accessibility under dynamic traffic conditions and flooding hazards with distinct drivers, and how accessibility loss is distributed across vulnerable communities within and across metropolitan regions. Specifically, our research questions are: (1) How does healthcare accessibility evolve under temporally varying traffic congestion? (2) How is healthcare accessibility impacted by flooding hazards? (3) What are drivers to such flood-induced accessibility loss? (4) How is flood-induced accessibility loss distributed across communities? Our analysis covers 30 metropolitan areas in the United States, spanning both inland and coastal urban settings. For each city, we use mesoscopic dynamic traffic simulation to represent diurnal traffic conditions and integrate these simulations with healthcare facility locations from OpenStreetMap and census block group population data to estimate dynamic healthcare facility accessibility. We then quantify accessibility loss under pluvial, fluvial and coastal flooding, as well as their combined effects (which we term as composite flooding), by accounting for road exposure, changes in traffic conditions on the perturbed network and the exposure of healthcare facility locations (Fig. \ref{fig:1} (c)). This design allows us to identify how different flood mechanisms alter healthcare access and to assess city-specific sensitivity to distinct flood footprints. We further examine how structural and functional drivers combine to shape city vulnerability. Finally, we evaluate how the burden of accessibility loss is distributed across communities through its spatial and social inequality.

Our findings show that hospital accessibility is shaped by the interaction of daily traffic dynamics, flood-specific disruption and the socio-spatial structure of urban systems. Across 30 metropolitan areas, congestion alone produces substantial temporal losses in access, reaching 28.8\% in Philadelphia, while pluvial, fluvial, coastal and composite flooding generate distinct and often much larger declines, ranging from 39.6\% in Boston under coastal flooding to 90.4\% in Houston under both pluvial and composite flooding. These impacts are governed not simply by inundation magnitude and extent, but by the combined effects of road exposure, fragility of hospital access, congestion amplification and hospital isolation induced by flooding. Grouping cities according to these mechanistic drivers explains accessibility loss more effectively than a single universal relationship and reveals contrasting structural and functional vulnerability profiles. Flood disruption also redistributes accessibility across neighborhoods, generally amplifying spatial inequality, with the largest increase in Houston ($\Delta$GI = 0.66 ), while its social burden varies across cities and flood types with Dallas showing strongest shift toward losses concentrated among socially vulnerable populations. Together, these results show that resilience of healthcare access cannot be inferred from congestion or flood exposure alone, but must be understood through the combined processes that determine when access is lost, where losses concentrate and which communities bear the greatest burden. By linking dynamic traffic conditions, multiple flood drivers and socio-spatial inequality in a cross-city analysis of hospital access, this study extends current approaches to urban accessibility and resilience and provides actionable insight for urban planners, transport agencies, healthcare providers and emergency managers in prioritizing corridor protection, emergency routing, hospital service planning and targeted resilience investment.

\section*{Results}\label{sec2}

\begin{figure}[h!] \centering \includegraphics[width=1\columnwidth,height=23cm]{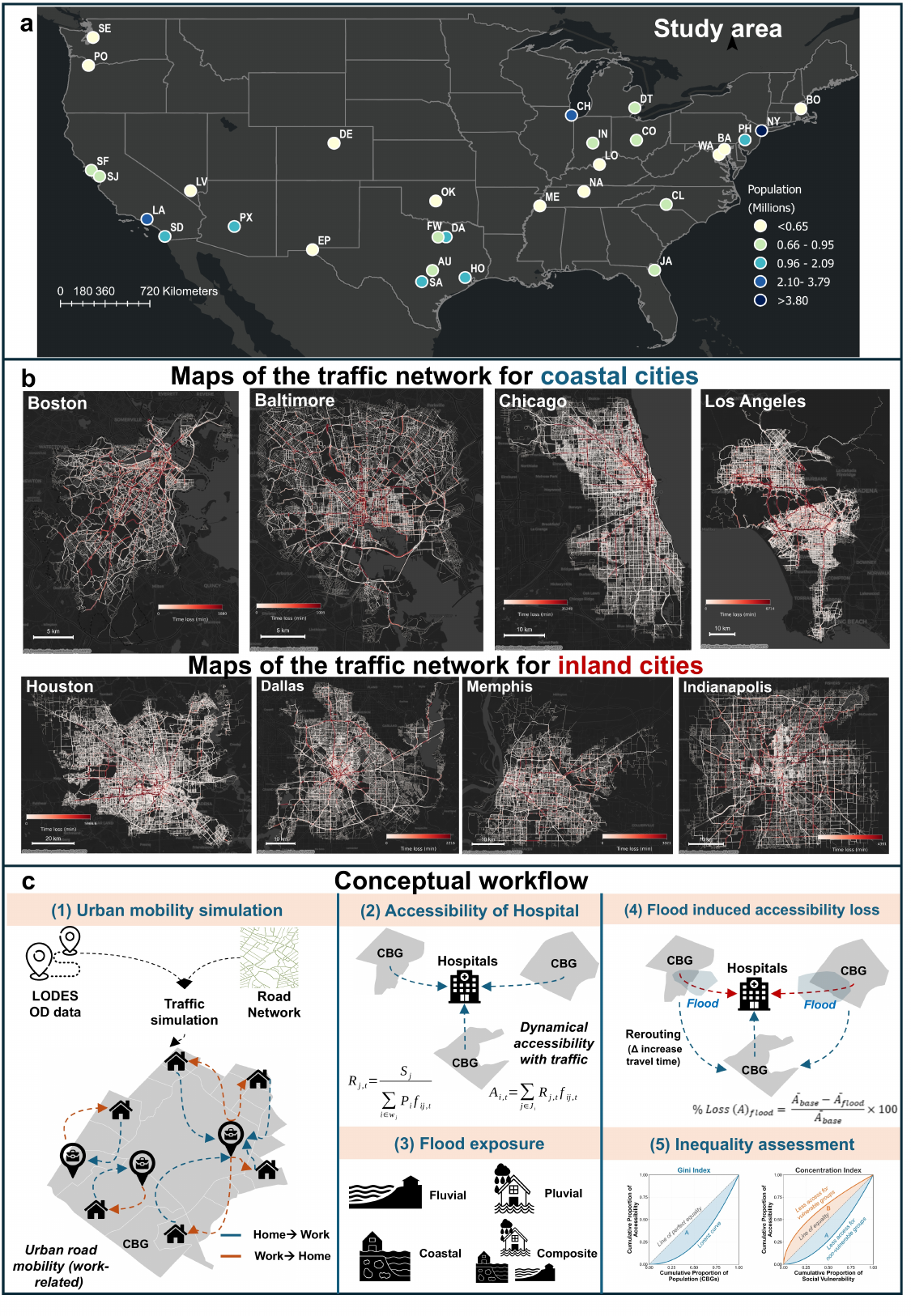}  \caption{}

\label{fig:1} \end{figure}

Our analysis covers 30 metropolitan areas across the contiguous United States, spanning a broad range of regional settings, urban forms and network structures (Fig. \ref{fig:1} (a)). The selected cities differ markedly in population, spatial extent and jurisdictional footprint. Some, such as Boston, Baltimore and San Francisco, are relatively compact, with dense road networks concentrated within constrained extents {(Fig. \ref{fig:1} (b) and Fig. S1)}. Others, including Houston, Dallas and Jacksonville, have larger and more spatially dispersed urban footprints (Fig. \ref{fig:1} (b) and Fig. S1). We further distinguish between coastal and inland cities to assess how accessibility disruption varies under different flood settings.

Urban mobility is represented using mesoscopic traffic simulation of work-related travel. Origin–destination flows are derived from the Longitudinal Employer-Household Dynamics Origin Destination Employment Statistics (LEHD–LODES) commuting data (\cite{uscensus_lodes}) (see Methods), which are widely used in urban mobility analysis (\cite{kasmalkar2020floods,xu2024unified}), to simulate city-wide traffic conditions over a 24-hour period spanning both peak and non-peak hours. These simulations provide a spatio-temporally resolved representation of network performance, allowing us to track time loss across road segments and to characterize realistic congestion propagation at city scale. Fig. \ref{fig:1} (b) illustrates these traffic patterns, with road segments colored from white to red to indicate increasing time loss under congestion. The simulated traffic patterns reveal pronounced heterogeneity both within and across cities, with distinct congestion structures emerging in coastal and inland systems. In coastal cities, congestion is typically concentrated along dense central corridors and a limited number of radial links, most clearly in Boston and Baltimore, where high-delay segments cluster within compact urban extents (Fig. \ref{fig:1} (b)). Chicago shows a strong corridor-based structure of elevated delay, whereas Los Angeles exhibits a more polycentric pattern distributed across multiple major arterials. By contrast, inland cities such as Houston, Dallas and Indianapolis display broader and more spatially extensive congestion footprints, with delay spread across larger portions of the network rather than confined to a narrow urban core (Fig. \ref{fig:1} (b)). The magnitude of congestion also varies substantially across cities, reflecting differences in travel demand and network structure. Chicago and New York exhibit some of the highest spatial mean time losses among the study cities, reaching approximately 90 h and 128 h, respectively, whereas Detroit and Portland show much lower values of about 1 h and 5 h (Fig. \ref{fig:1} (b) and Fig. S1).

\subsection*{Traffic congestion induced dynamical accessibility loss}

\begin{figure}[h!] \centering \includegraphics[width=1\columnwidth,height=20 cm]{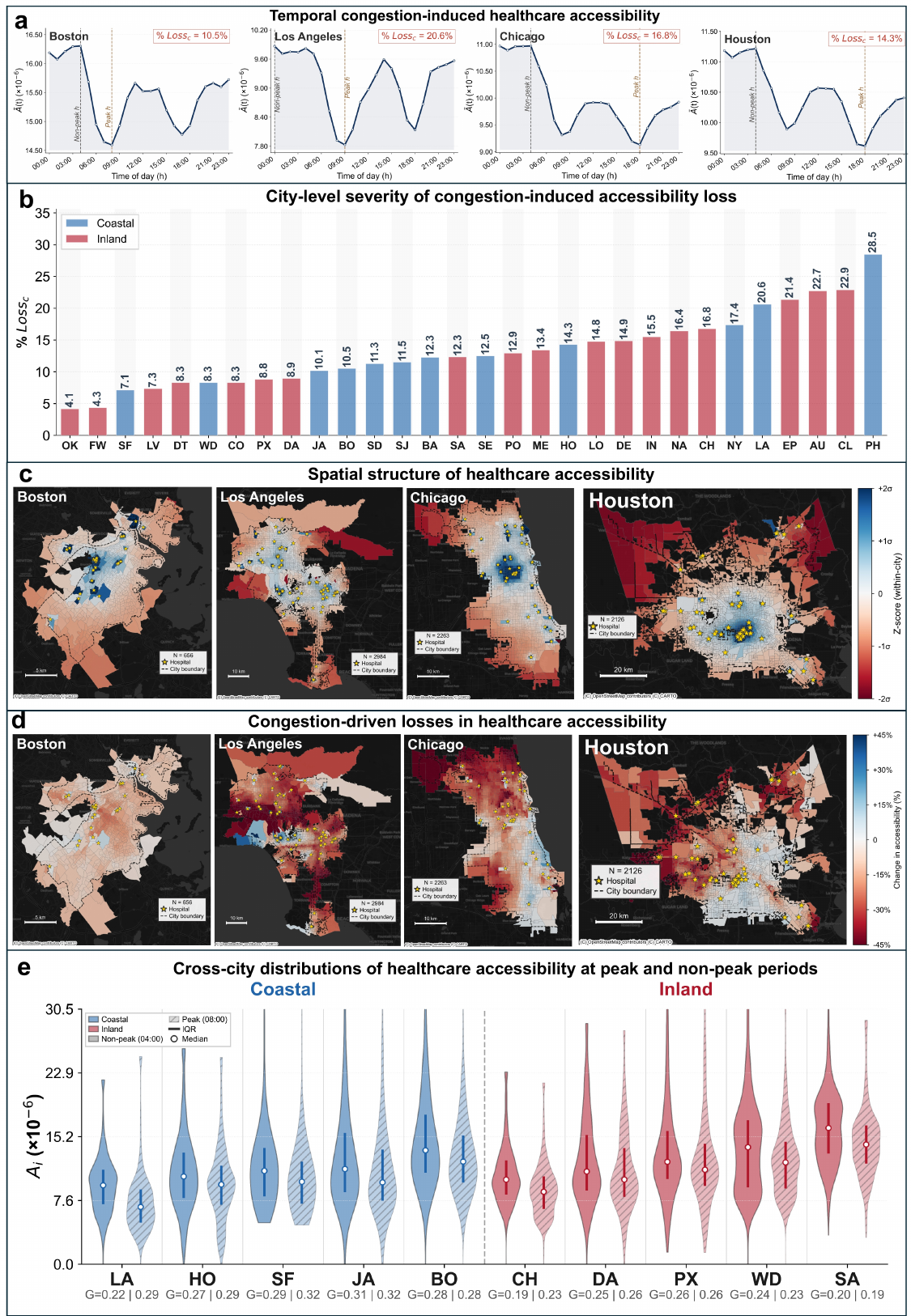} \caption{}

\label{fig:2} \end{figure}

Our population-weighted accessibility measure (M2SFCA) integrates hospital supply, census block group population demand and shortest-path travel times under dynamically varying traffic conditions; see Methods. Using traffic states derived from simulation, we evaluate hospital accessibility over a 24-hour period to capture its temporal variation. As shown in Fig. \ref{fig:2} (a) for four representative cities, Boston, Los Angeles, Chicago and Houston, accessibility follows a clear diurnal cycle, with highest values during late-night and early-morning hours and sharp declines during the peak commuting periods of 07:00 to 10:00 and 17:00 to 20:00. A pronounced morning trough is evident across all four cities, showing that recurrent rush-hour congestion alone can substantially reduce reachable hospital opportunities even in the absence of external disruption. This decline reflects the concentration of traffic on major commuter corridors, which increases travel times and reduces the number of hospitals reachable within the defined travel-time threshold.

The amplitude of this daily fluctuation differs across cities. Los Angeles shows the largest decline in accessibility between high- and low-access periods (\%$Loss_c$ = 20.6), followed by Chicago (16.8), Houston (14.3) and Boston (10.5) (Fig. \ref{fig:2} (a)). These differences indicate that the accessibility effects of congestion depend not only on traffic intensity, but also on network structure, hospital distribution and the extent to which populations rely on a limited set of high-burden corridors. The same broad temporal pattern is observed across the full set of cities, although the value of dip in accessibility varies by city. Austin, Chicago, Detroit, Houston, Las Vegas and Philadelphia exhibit greater accessibility loss during the evening peak, whereas most other cities show a deeper morning decline (Fig. S2). City-level losses span a broad range, from 4.1\% in Oklahoma City and 4.3\% in Fort Worth to 28.8\% in Philadelphia, the highest observed value, followed by Charlotte at 22.9\% (Fig. \ref{fig:2} (b)). These results show that routine congestion is not a uniform background condition, but a major source of temporal variation in hospital accessibility across cities.

In addition to these temporal dynamics, hospital accessibility exhibits substantial spatial variation across census block groups within cities. To enable cross-city comparison, we standardize accessibility using z-scores, thereby reducing bias arising from differences in population demand and hospital supply. We focus on non-peak conditions to isolate underlying spatial disparities in access, when congestion effects are comparatively limited. Fig. \ref{fig:2} (c) shows these patterns for Boston, Los Angeles, Chicago and Houston, with corresponding maps for the remaining cities provided in (Fig. S3). In all cities, accessibility is distributed unevenly across urban space, with higher values concentrated in central areas and around clusters of hospitals, and lower values in peripheral zones. This produces a consistent center–periphery gradient, although its spatial pattern varies across cities, indicating that accessibility is structured by distinct urban geographies rather than distributed uniformly across populations.

These intra-city patterns also differ systematically among cities. Boston and Chicago exhibit relatively compact high-accessibility cores with steep declines toward the periphery, indicating that effective hospital access is concentrated around dense central hospital clusters and diminishes rapidly with distance from these service centres. Los Angeles, by contrast, shows a more fragmented and polycentric pattern, with multiple localized high-access zones distributed across the urban area rather than organized around a single dominant core, consistent with its dispersed urban form and multiple activity centres. Houston combines a broad central accessible zone with an extensive low-accessibility periphery, suggesting that although hospitals are concentrated within the inner urban area, the city’s large spatial extent leaves peripheral census block groups with substantially lower relative accessibility (Fig. \ref{fig:2} (c)). The corresponding spatial patterns for the remaining cities are shown in (Fig. S3). These patterns indicate that intra-city accessibility is shaped jointly by the spatial concentration of hospitals and the road-network connectivity linking populations to those facilities.

To quantify how peak-hour congestion reshapes these spatial patterns, we compute the percentage change in accessibility for each census block group relative to non-peak conditions. Fig. \ref{fig:2} (d) shows predominantly negative impacts, but with distinct intra-urban structures. In Boston, losses are concentrated in central and inner urban areas, whereas Los Angeles exhibits a broader and more spatially uneven decline. Chicago shows a contrasting pattern, with substantial losses in western and northwestern areas but relative gains in parts of the eastern and southeastern corridor, while Houston is characterized by widespread losses across periphery and intermediate zones, with smaller declines or localized gains in parts of the urban core. The magnitude of this intra-urban variation also differs across cities, with Los Angeles and Chicago showing changes exceeding 35\%, whereas Boston remains below 15\%. Although most neighborhoods experience accessibility decline, small pockets of positive change emerge in several cities. This reflects a competition effect: as congestion reduces effective access for some neighborhoods, demand pressure on shared hospital destinations also declines, allowing certain better-connected locations to become relatively more accessible (Fig. \ref{fig:2} (d) and Fig. S4). Peak-hour congestion therefore does not reduce accessibility uniformly across urban space, but redistributes it by imposing stronger penalties on some neighborhoods than on others. 

This redistribution is also evident in the inter-city accessibility distributions shown in (Fig. \ref{fig:2} (e)). In most cities, peak-hour accessibility shifts downward and the Gini coefficient increases, indicating that congestion widens inequality between better- and worse-served census block groups. The increase is especially pronounced in several coastal cities, including Los Angeles (0.223 to 0.293), San Francisco (0.295 to 0.319) and Jacksonville (0.307 to 0.319), consistent with their more corridor-concentrated accessibility structures. By contrast, some inland cities show little change in GINI value between non-peak and peak conditions. In San Antonio, the distributions are nearly unchanged and inequality remains stable, whereas Washington, D.C. shows a slight decline in Gini under peak-hour conditions, suggesting that more spatially distributed access patterns can moderate the inequality effects of congestion. The corresponding distributions for all other cities are provided in (Fig. S5).  Overall, these findings show that congestion not only lowers hospital accessibility, but also amplifies existing intra-urban inequality, with the strongest distributional effects occurring where access is concentrated along a limited set of urban corridors.

\subsection*{Flood exposure and road-network disruption} 

To quantify external disruption to healthcare accessibility, we consider pluvial, fluvial and coastal flooding, induced respectively by extreme precipitation, upstream river discharge and elevated tide levels. Flood exposure is characterized by overlaying road networks with 100-year return-period inundation rasters and summarizing both total flood volume within each city and the share of road length inundated. Dynamic road closures are then used to evaluate how flood-induced capacity loss propagates through the traffic network (see Methods).

\begin{figure}[h!] \centering \includegraphics[width=1\columnwidth]{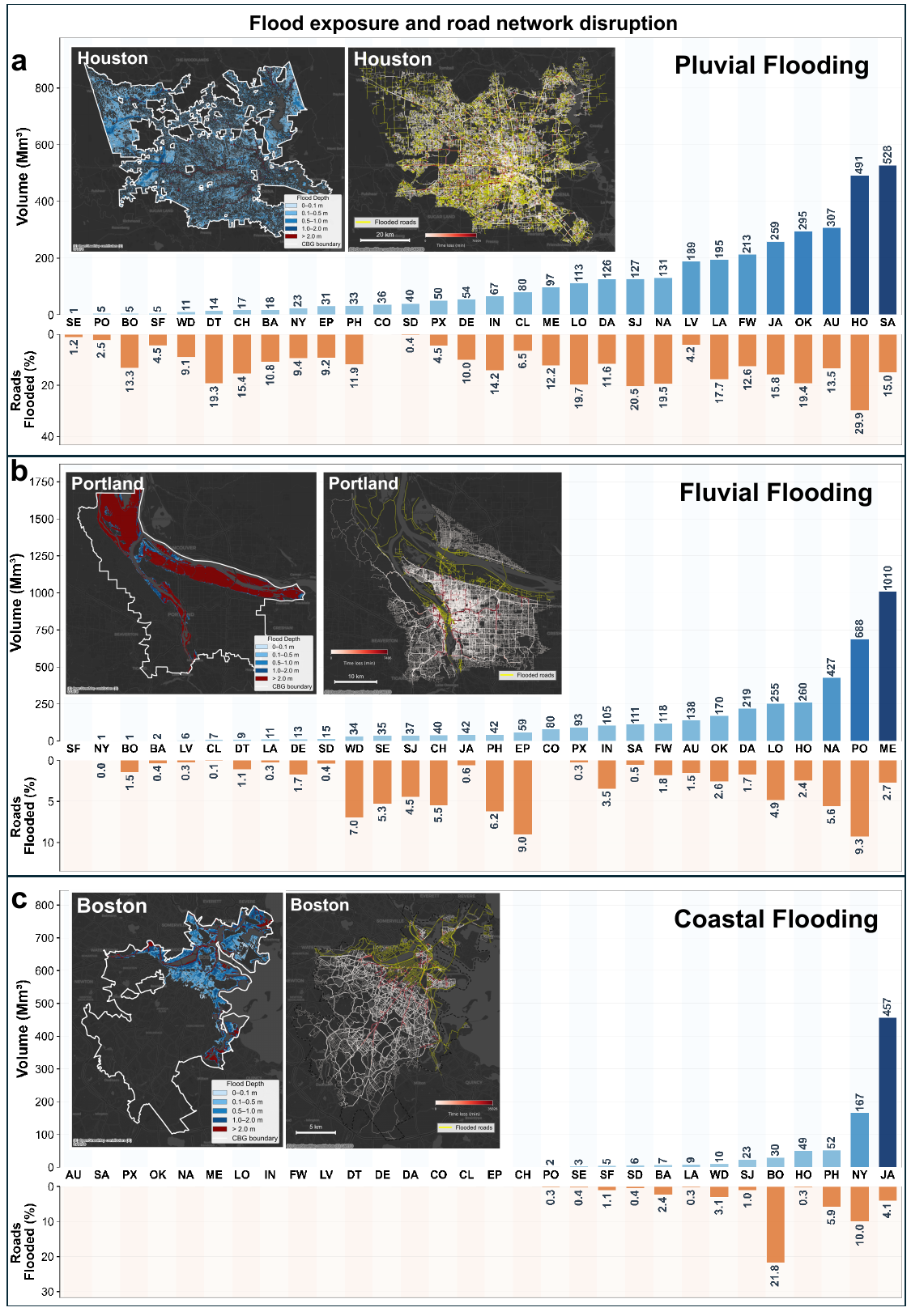} 
\caption{}

\label{fig:3} \end{figure}

Flood exposure varies sharply across drivers and cities (Fig. \ref{fig:3}). Under pluvial flooding, inundation volume ranges from negligible levels in Seattle and Portland to more than 500 Mm$^{3}$ in San Antonio, yet road disruption does not scale directly with flood volume. Houston shows the highest flooded-road share (29.9\%), despite lower total inundation than San Antonio, and experiences the increased traffic delay, with maximum time loss increasing up to 14-fold (Fig. \ref{fig:3} (a)). Under fluvial flooding, the largest inundation volumes occur in Memphis, Portland and Nashville, whereas Portland, El Paso and Washington, D.C. show relatively high road exposure compared with more heavily inundated cities; in Portland, mean travel delay increases by about 10\% (Fig. \ref{fig:3} (b)). Coastal flooding is confined to coastal cities, with Jacksonville showing the highest inundation volume and Boston the largest flooded-road share; in Boston, mean travel delay increases by 11.2\% (Fig. \ref{fig:3} (c)). These contrasts show that flood magnitude alone is a poor predictor of road-network disruption.To quantify the combined effect of these hazards, we merge pluvial, fluvial and coastal inundation into a composite flood layer (see Methods; Fig. S6). Under this combined scenario, Houston again shows the strongest traffic response, with mean time loss increasing by nearly threefold. The Transport impacts depend on where flood waters intersect the road system and how the loss of flooded links redistributes traffic across the remaining network.

\subsection*{Flood induced dynamical accessibility loss}

\begin{figure}[h!] \centering \includegraphics[width=1\columnwidth]{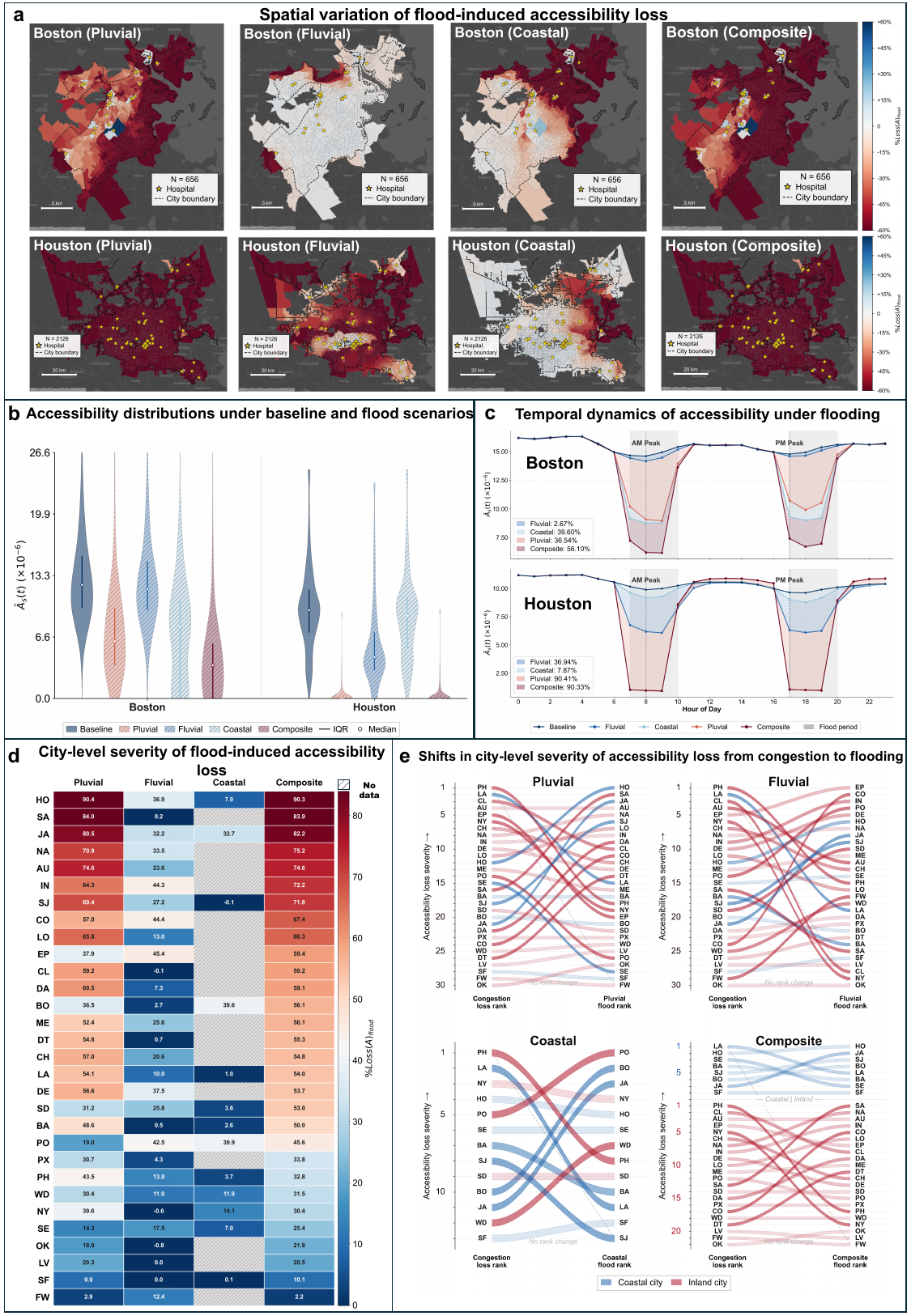} 
\caption{}

\label{fig:4} \end{figure}

Flood-induced accessibility loss varies strongly across flood drivers and between cities (Fig. \ref{fig:4} (a) and Fig. S7-S10)). Without loss of generality, we highlight two representative cities, Boston and Houston for discussion. In Boston, accessibility losses remain spatially differentiated across hazard types. During peak-hour conditions, pluvial flooding produces the largest decline in mean accessibility, concentrated in the eastern part of the city (48.36\%), whereas fluvial impacts are much smaller and remain localized along river corridors (3.59\%). Coastal flooding also generates substantial accessibility loss (43.10\%), concentrated mainly in northern Boston. Under composite flooding, mean accessibility declines by 67.63\%, and the spatial footprint of loss extends across much of the city. Houston, by contrast, shows a broader and more severe disruption pattern. Under pluvial flooding, accessibility declines across much of the urban area and reaches its largest reduction (95.84\%), whereas fluvial (43.91\%) and coastal (8.61\%) impacts remain more spatially confined to census block groups near river channels and the coast. The composite scenario shows a pattern similar to pluvial flooding, with mean accessibility loss reaching 95.8\% during peak traffic conditions. Small localized gains are also visible in both cities, consistent with a competition effect in which flood-induced disconnection of some neighborhoods reduces demand pressure on shared hospital opportunities in better-connected areas (Fig. \ref{fig:4} (a) and Fig. S7-S10). Fig. \ref{fig:4} (b) further shows that flooding shifts the accessibility distributions downward and, in some cities, broadens their spread, indicating that losses are not uniform across census block groups; this pattern is especially evident in Houston, where widespread flooding produces a pronounced downward shift in mean accessibility under all hazard scenarios. These patterns also observed across cities (Fig. S11) which indicates that flood impacts on hospital accessibility are shaped not only by hazard extent, but also by hospital distribution, road-network structure and pre-existing spatial inequality interacts with flood hazard.

Flooding reduces citywide hospital accessibility during the hours of inundation, but the magnitude of decline varies strongly across flood mechanism (Fig. \ref{fig:4} (c) and Fig. S12)). In Boston, coastal flooding produces the largest loss (39.6\%), followed by pluvial flooding (36.54\%) and fluvial flooding (2.67\%), whereas composite flooding yields the greatest overall decline (56.10\%) relative to the baseline no-flood condition. Houston shows a much stronger response and a different hazard ordering: pluvial flooding produces the largest accessibility loss (90.41\%), followed by fluvial flooding (36.94\%), whereas coastal flooding has a comparatively limited effect (7.87\%); composite flooding reduces accessibility by 90.33\%. Similar hazard-specific temporal declines are observed across the full set of cities, although their magnitude varies substantially by flood driver (Fig. S12). These trajectories show that flood impacts on hospital accessibility are dynamic functional disruptions that evolve over time, rather than static reductions in access.

We next summarize accessibility loss across all cities and flood mechanisms; see Fig. \ref{fig:4} (d). The cities are arranged in descending order based on pluvial flood-induced loss. Accessibility loss differs widely across both hazard type and urban system, with composite flooding generally producing the largest declines. The strongest pluvial impacts occur in Houston (90.4\%) and San Antonio (84.0\%), whereas Fort Worth (2.9\%) and San Francisco (9.9\%) show only limited change. Fluvial flooding is most severe in El Paso (45.4\%) and Columbus (44.4\%), while Charlotte, San Francisco and Las Vegas show little to no loss. Boston experiences the largest coastal accessibility decline (39.6\%), whereas San Jose shows almost no coastal effect (-0.1\%). Composite flooding produces the greatest overall losses, particularly in Houston (90.3\%) and Jacksonville (82.2\%) among coastal cities, and in San Antonio (83.9\%) and Nashville (75.2\%) among inland cities.  

To evaluate how flood exposure alters city-level severity of accessibility loss relative to unperturbed traffic conditions, we compare city severity rankings under baseline congestion and under each flood scenario, (where a lower rank indicates greater severity) (Fig. \ref{fig:4} (e)). City severity under flooding does not simply mirror severity under routine congestion; instead, rankings shift substantially once road networks are exposed to flood hazards. Pluvial flooding produces the largest overall reordering, with Houston moving from a moderate congestion-related loss (14.3\%) to extremely high losses under pluvial (90.4\%) and composite flooding (90.3\%), placing it among the most severe cities under flood conditions. Jacksonville shows the largest increase in severity rank under pluvial flooding (+18), whereas Columbus exhibits the largest shift under fluvial flooding (+22), indicating that riverine flooding can sharply amplify accessibility loss even in cities with relatively modest congestion impacts. Coastal flooding reveals a different pattern: Los Angeles has one of the highest congestion losses (23.2\%) but only a limited coastal flood amplification (1.0\%), whereas Boston, despite a lower congestion loss (10.5\%), experiences much stronger coastal-flood amplification (39.6\%). This reordering occurs because flood impacts depend not only on baseline congestion, but also on which road segments are inundated, how much network redundancy is available, where hospitals are located and how rerouting redistributes traffic through the remaining network.

\subsection*{Mechanistic Drivers of flood-induced accessibility loss and city vulnerability typologies}

\begin{figure}[h!] \centering \includegraphics[width=1\columnwidth]{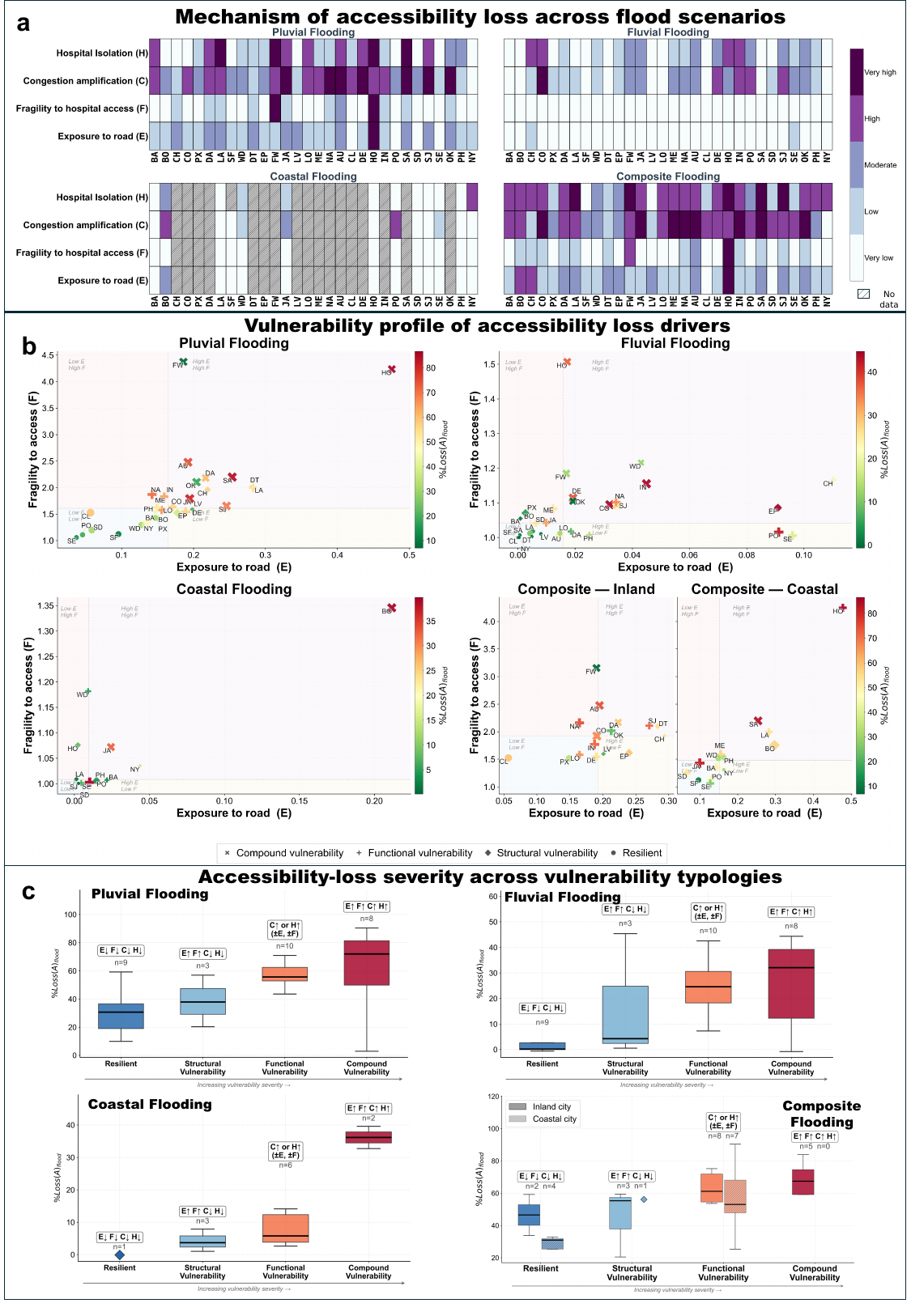} 
\caption{}

\label{fig:5} \end{figure}

Flood-induced accessibility loss varies widely across cities and cannot be explained by flood exposure alone. In some cases, cities with comparatively lower flood exposure, such as Jacksonville and Austin, experience substantial accessibility loss, whereas cities such as Detroit and Oklahoma City show higher exposure but much smaller reductions in access (Fig. \ref{fig:3} (a),  Fig. \ref{fig:4} (d)). This contrast indicates that city vulnerability depends not only on flood extent, but also on how hazards interact with road-network structure, hospital configuration and traffic response. To understand why similar levels of flood exposure produce different accessibility outcomes, we quantify the structural and functional drivers of accessibility loss. Structural drivers include road-network exposure (E) and fragility of hospital access (F), whereas functional drivers include congestion amplification (C) and hospital isolation (H) under flood-perturbed conditions (see Methods). Fig. \ref{fig:5} (a) shows the standardized values of these four drivers across pluvial, fluvial, coastal and composite flooding. Higher standardized values indicate stronger road exposure, greater fragility of hospital access, larger congestion amplification or more severe hospital isolation in a given city under the corresponding flood scenario. Driver values vary substantially across cities and flood types. Under pluvial flooding, Houston exhibits high values across all four dimensions, whereas Seattle remains low across most drivers. Under fluvial flooding, structural drivers are generally weaker than under pluvial flooding, but functional drivers remain elevated in several cities, indicating that even limited direct exposure can still generate substantial accessibility disruption through congestion and hospital isolation. Under coastal flooding, Boston stands out with high exposure, fragility and hospital isolation relative to other coastal cities. Composite flooding produces the most widespread multi-driver burden, with elevated values across several dimensions in many cities. Overall, structural drivers vary more strongly with the spatial extent of inundation, whereas functional drivers remain comparatively persistent across flood scenarios.

Building on the driver patterns in Fig. \ref{fig:5} (a), we next examine how cities organize into vulnerability typologies in driver space ({Fig. \ref{fig:5} (b) and Fig. S13}). Cities separate into distinct groups rather than following a single linear relationship between mechanistic drivers and accessibility loss. Under pluvial flooding, high-loss cities cluster where both exposure and fragility are elevated, with Houston occupying the most extreme position and exhibiting the largest accessibility loss. Cities with low exposure and low fragility remain comparatively resilient, whereas intermediate cities fall into structural and functional vulnerability groups depending on whether loss is associated primarily with elevated exposure–fragility conditions or with stronger congestion amplification and hospital isolation; the latter driver space is shown in ({Fig. S12}). Under fluvial flooding, the exposure range is narrower, but cities with modest exposure and higher fragility still show substantial loss, indicating that even limited direct inundation can produce marked accessibility decline when hospital access is highly sensitive to disruption. This is evident in Houston, while cities such as Columbus and Indianapolis also show high accessibility loss despite occupying a more moderate exposure–fragility space, because their congestion amplification and hospital isolation are substantially stronger ({Fig. S13}). Coastal flooding shows the clearest separation, with Boston standing apart from other coastal cities because relatively high exposure coincides with elevated fragility, stronger congestion amplification and greater hospital isolation, together producing high accessibility loss. Under composite flooding, inland and coastal cities form distinct clusters, with the highest losses occurring where multiple drivers are simultaneously elevated. These patterns show that city vulnerability is multi-driver in nature, and that accessibility loss is shaped by the joint configuration of structural and functional stressors rather than by exposure alone.

To examine whether these vulnerability profiles correspond to systematic differences in accessibility-loss severity, we compare losses across four typologies defined from the joint configuration of mechanistic drivers relative to median thresholds: resilient, structural vulnerability, functional vulnerability and compound vulnerability (see methods; {Fig. \ref{fig:5} (c)}). A clear gradient emerges across flood mechanisms, with accessibility loss increasing from resilient to compound-vulnerability cities. Under pluvial flooding, losses rise progressively across classes, with resilient cities such as Boston and San Francisco showing relatively small declines, whereas compound-vulnerability cities such as Houston and Jacksonville exhibit the largest losses. The corresponding city assignments for each typology are listed in ({Table S2}). A similar ordering is evident under fluvial flooding, where resilient cities, including Los Angeles and Las Vegas, experience only minor loss, while functional-vulnerability cities such as Chicago and Austin, and compound-vulnerability cities such as San Antonio and Houston, show substantially greater decline. Coastal flooding shows the strongest separation among classes, with lower-severity groups experiencing only modest accessibility loss, whereas compound-vulnerability cities stand apart with much larger reductions, most clearly in Boston, where all mechanistic drivers are elevated. Composite flooding preserves the same overall pattern across both inland and coastal subsets, with the greatest losses again concentrated in the most vulnerable classes.  Here we notice that cities with high functional vulnerability losses more access compared to cities with structural vulnerability across the flood mechanisms. Taken together, this typology shows that accessibility loss does not follow a single universal law, but instead emerges from different combinations of exposure, fragility, congestion amplification and hospital isolation across cities.

\subsection*{Spatial and social inequality of access}
\begin{figure}[h!] \centering \includegraphics[width=1\columnwidth]{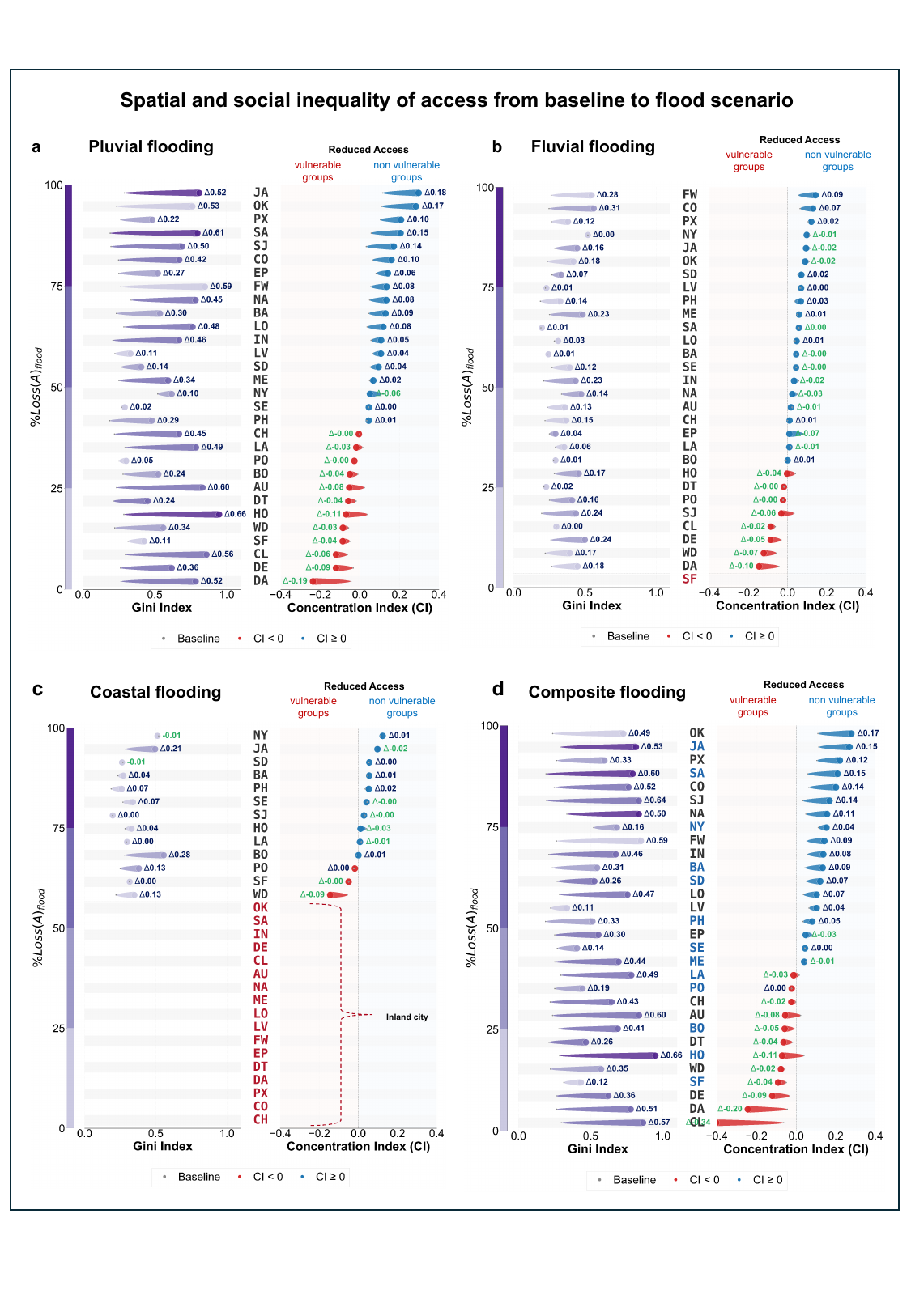} 
\caption{}

\label{fig:6} \end{figure}

To quantify flood-induced changes in the spatial distribution of accessibility, we measure the population-weighted Gini index (GI) for each city under all flood scenarios and compare it with the baseline condition ({Fig. \ref{fig:6}}). Flooding generally amplifies spatial inequality in hospital accessibility, although the magnitude of change varies across hazard types and cities. The largest shifts occur under pluvial and composite flooding, indicating that widespread network disruption tends to intensify within-city disparities more strongly than fluvial or coastal flooding. Under pluvial flooding, the greatest increases in GI are observed in Houston ($\Delta$GI = 0.66), San Antonio (0.61), Austin (0.60) and Fort Worth (0.59) (situated in state of Texas), showing that accessibility loss is concentrated unevenly across census block groups rather than distributed uniformly across the urban system ({Fig. \ref{fig:6} (a)}). Composite flooding shows a similarly strong pattern, with large increases in Houston (0.66), San Jose (0.64), San Antonio (0.60) and Nashville (0.50) ({Fig. \ref{fig:6} (d)}). By contrast, fluvial flooding produces smaller and less consistent inequality amplification, and high accessibility loss does not always coincide with large increases in GI. For example, El Paso and Portland experience substantial fluvial accessibility loss but comparatively limited amplification of spatial inequality, indicating that riverine flooding can reduce access broadly without necessarily sharpening its within-city concentration ({Fig. \ref{fig:6} (b)}). Coastal flooding is restricted to coastal cities and shows its strongest inequality amplification in Boston ($\Delta$ GI = 0.28) ({Fig. \ref{fig:6} (c)}).

A remaining question is whether flood-induced accessibility loss is distributed evenly across vulnerable and non-vulnerable communities. To assess this, we use the concentration index (CI) to quantify the social gradient of accessibility loss. In contrast to the Gini index, which generally increases under flooding, CI shifts in both directions across cities and flood types, indicating that the social burden of accessibility loss is more heterogeneous ({Fig. \ref{fig:6}}). Under pluvial flooding, Dallas (baseline CI = -0.045), Denver (-0.031) and Charlotte (-0.063) already show baseline inequality disadvantaging socially vulnerable populations, and this pattern intensifies further under flooding, with $\Delta$ CI values of -0.19, -0.09 and -0.06, respectively. These shifts indicate that communities already facing weaker access become further disadvantaged under disruption. By contrast, Jacksonville ($\Delta$ CI = 0.18) and Oklahoma ($\Delta$ CI = 0.17) shift in the opposite direction, indicating relatively greater losses among less vulnerable groups ({Fig. \ref{fig:6} (a)}). A similar bidirectional pattern appears under composite flooding: Oklahoma ($\Delta$ CI= 0.17), Jacksonville ($\Delta$ CI= 0.15) and Phoenix ($\Delta$ CI= 0.12) show positive shifts, whereas Charlotte ($\Delta$ CI= -0.34), Dallas ($\Delta$ CI= -0.122) and Denver ($\Delta$ CI= -0.09) show strongly negative shifts ({Fig. \ref{fig:6} (d)}). Fluvial flooding generates smaller CI responses overall, with the largest positive shift in Fort Worth ($\Delta$ CI= 0.09) and the strongest negative shift in Dallas ($\Delta$ CI= -0.10) ({Fig. \ref{fig:6} (b)}), while many cities still show a shift toward greater losses among vulnerable populations. Coastal flooding, by contrast, shows uniformly positive but generally weak CI values suggesting that accessibility loss is more concentrated towards the less vulnerable community ({Fig. \ref{fig:6} (c)}) consistent with the tendency for affluent households in many coastal cities to be concentrated closer to the coast \cite{smith2020coastal}. These patterns show that, Unlike spatial inequality, which generally increases under flooding, the social burden of accessibility loss depends on the local alignment between hazard exposure, healthcare infrastructure and residential vulnerability.

These findings show that assessing hospital accessibility requires accounting for both temporal traffic dynamics and the shifting effects of distinct and composite flood hazards. City resilience to critical care access is shaped not only by the magnitude of accessibility loss, but also by when that loss occurs, how it propagates through the transport network, and which communities bear the greatest burden. Measuring accessibility decline alone is therefore insufficient; understanding its spatial and social distribution is essential for identifying the resilience of urban healthcare accessibility.

\section*{Discussion}
With nearly two thirds of the world’s population projected to live in cities by 2050 (\cite{un2018urbanization}), sustaining access to critical facilities is becoming an increasingly important dimension of urban resilience (\cite{colglazier2015sustainable}). Healthcare is especially consequential because delays in reaching hospitals can directly affect emergency response, treatment and community well-being (\cite{gulliford2002does,salas2024impact}). Although the concept of the “x-minute city” has drawn attention to threshold-based access to essential services, accessibility is still commonly assessed using static metrics or within the context of a single disruption type or a single region (\cite{bruno2024universal,yu2020disruption,he2021flood}). Our results show that such perspectives miss an important feature of healthcare access under disruption. When the connecting transport infrastructure is perturbed by flooding, changes in accessibility are not determined by flood exposure alone, but also by the interaction of temporally varying traffic, distinct flood processes and the spatial distribution of roads, hospitals and population demand. By combining mesoscopic traffic simulation with pluvial, fluvial, coastal and composite flood footprints, we show that flood-induced accessibility loss reflects not only direct exposure of roads and facilities, but also spillover effects that propagate through the wider transport network and redistribute losses across neighborhoods.

Our findings show that accounting for traffic dynamics substantially changes the assessment of healthcare accessibility, even in the absence of external disruption. Across the study cities, peak-hour congestion alone reduces hospital accessibility from 4.1\% (in Oklahoma) up to 28.5\% (in Philadelphia), while also widening inequalities in access, particularly in coastal cities where congestion is more strongly concentrated along a limited set of urban corridors. This pattern indicate that congestion is not simply a baseline mobility condition, but a persistent constraint on healthcare access that shapes temporal loss and intra-urban inequality even in the absence of external disruption. Flooding compounds these constraints, but the severity and form of disruption differ sharply across cities and underlying flood hazard mechanisms. Under pluvial flooding, accessibility loss ranges from 2.9\% in Fort Worth to 90.4\% in Houston; under fluvial flooding, from 0.1\% in Charlotte to 45.4\% in El Paso; and under coastal flooding, from 0.1\% in San Francisco to 39.6\% in Boston. Composite flooding produces the largest overall declines, indicating that concurrent flood processes create the strongest constraints on transport-dependent healthcare access. These contrasts show that accessibility loss is governed not only by direct flood exposure of roads and healthcare infrastructure, but also by how specific hazards interact with road-network structure, hospital locations and pre-existing traffic conditions. As a result, disruption can propagate through the wider network and generate substantial accessibility losses even in areas distant from flood zones, so that cities with comparable flood exposure can experience very different outcomes. 

Beyond average accessibility loss, flooding also alters its spatial and social distribution of accessibility to healthcare. Flooding generally amplifies spatial inequality in accessibility, especially under pluvial and composite flooding, indicating that disruption is concentrated unevenly across neighborhoods rather than shared uniformly across the urban system. The social distribution of accessibility loss is more variable across cities. In some cases, losses fall more heavily on socially vulnerable communities, whereas in others they are concentrated among less vulnerable populations, reflecting how flood-exposed roads and hospitals overlap with the residential geography of vulnerability. Coastal flooding, in particular, more often shifts losses toward less vulnerable communities because coastal exposure in many cities overlaps with higher-income neighborhoods and the access corridors on which they rely (\cite{smith2020coastal}), whereas pluvial and composite flooding produce more mixed accessibility loss among socially vulnerable communities. Overall, our study shows that resilience of healthcare access is defined not only by how much accessibility is lost, but also by its timing, spatial concentration and social distribution.

Why cities with apparently similar flood exposure experience very different accessibility losses becomes clearer when accessibility loss is examined through structural and functional drivers. Road-network exposure and fragility of hospital access help explain the structural sensitivity of cities to flooding, whereas congestion amplification and hospital isolation capture how disruption propagates through the transport–healthcare system. Under free flow (no flooding) conditions, accessibility declines with travel time in a relatively consistent power-law form across cities (\cite{zhong2025universal}), but this pattern breaks down once flooding and traffic dynamics are introduced. Accessibility loss then reflects different combinations of local exposure, rerouting, hospital isolation, congestion amplification and network fragility, all of which vary across cities and neighborhoods. Flood-disrupted accessibility does not conform to a single scaling relationship across cities; grouping cities by structural and functional drivers therefore provides a clearer explanation of loss than flood extent alone. In particular, losses increase monotonically from resilient cities, where all drivers remain below threshold, to compound-vulnerability cities, where all four drivers are elevated. For urban sustainability and resilience planning, this suggests that reducing flood exposure alone is insufficient if functional disruption within the transport network remains unresolved as we observed in Columbus and Indianapolis.

This study has several limitations that point to directions for future research. First, the traffic simulations are based on LODES origin–destination data (\cite{uscensus_lodes}), which primarily represent employer–household commuting flows and therefore do not capture the full diversity of urban travel beyond work-related trips. Although this focus is appropriate for representing major daily traffic dynamics, broader mobility datasets, including those derived from mobile phone records (\cite{ambuhl2023understanding,fan2022equality}), could improve representation of non-work travel and provide a more complete assessment of accessibility. Relatedly, the present analysis considers only private vehicle travel (\cite{dave2025local}). While access to healthcare often depends on both public and private modes, private vehicles dominate travel in many US metropolitan areas and also provide critical last-mile connectivity (\cite{census_commuting_acs_2025}). Future work could therefore incorporate multimodal transport networks and individual mobility behaviour to strengthen the robustness of accessibility estimates. Second, flood exposure is represented using the maximum inundation footprint rather than the temporal evolution of flood extent (\cite{kasmalkar2020floods,martello2021evaluation}). As a result, the analysis captures the spatial reach of hazard exposure but not the timing and duration through which disruption unfolds. Incorporating time-varying inundation dynamics would allow future studies to represent more realistic sequences of infrastructure failure and recovery. Third, owing to the lack of detailed hospital capacity data, all hospitals are treated as having equal capacity in the accessibility analysis (\cite{fang2025evaluating}). This assumption simplifies the supply side of healthcare access and may obscure differences in the ability of facilities to absorb demand during disruption. Including information on hospital size, service capacity and surge capability would improve representation of healthcare supply and refine interpretation of accessibility loss. The future studies can also expand the understanding by incorporating other critical facilities such as shelter locations and transportation hubs and propel the reseacrh towards optimizing the critical facilities not only in ideal condition but also under external disruption. Despite these limitations, the study provides a multi-city account of how distinct flood hazards reshape healthcare accessibility through interacting structural and functional mechanisms. By identifying when accessibility is lost, where losses concentrate and how they are distributed across communities, it offers evidence that can inform more precise and equity-oriented urban planning, disaster preparedness and resilience policy. 

In conclusion, this study shows that protecting healthcare access in cities requires planning beyond hospitals location alone. Access to care depends on both the physical exposure of roads and hospitals and the way disruption spreads through the wider transport network, often extending losses far beyond the flooded zone. This has direct implications for disaster management and evacuation planning, where preserving movement along a limited set of critical access corridors may be as important as protecting hospital buildings themselves. The results also provide a basis for future infrastructure planning by identifying which cities are most sensitive to pluvial, fluvial, coastal or compound flooding, and by showing whether losses are driven mainly by direct road exposure, weak network redundancy, congestion amplification or hospital isolation. Such information can help planners prioritize flood protection for the hazard types that pose the greatest threat in each city, strengthen emergency routing and backup access, and guide decisions about where additional healthcare capacity or future hospital locations would improve coverage and reduce vulnerability. More broadly, the study argues for a system-level approach to urban resilience in which reliable and equitable access to essential care is treated as a core planning objective under both everyday congestion and climate-driven disruption.

\section*{Method}
\subsection*{Study area}

This study focuses on the 30 most populous cities in the United States, identified using the 2022 U.S. Census (\cite{uscensus}). Population was used as the selection criterion to avoid bias toward either the most traffic-congested or the most flood-exposed cities, as the objective was to jointly examine both dimensions in shaping accessibility to critical facilities. Collectively, these 30 cities account for $\approx$ 41.44 million residents, representing approximately 7\% of the total U.S. population and xx USD as GDP. The selected cities span 21 U.S. states, with multiple cities located in Texas (six), California (four), and Tennessee (two) ({Fig. \ref{fig:1} (a)}).

Based on their hydrologic setting, the cities were classified as coastal or inland. Coastal cities were defined as those located along the coastline or influenced by tidal processes and sea-level rise. The coastal cities group includes Baltimore, MD (Patapsco River); Boston, MA (Charles River); Houston, TX (Buffalo Bayou); Jacksonville, FL (St. Johns River); Los Angeles, CA (Los Angeles River); New York, NY (Hudson River); Philadelphia, PA (Delaware River); Portland, OR (Willamette River); San Diego, CA (San Diego River); San Francisco, CA; San Jose, CA (Guadalupe River); Seattle, WA (Duwamish River); and Washington, DC (Potomac River). Inland cities were defined as those not directly connected to the coast but closely associated with major river systems or river crossings that shape local flood dynamics. This group includes Austin, TX (Colorado River); Charlotte, NC (Catawba River); Chicago, IL (Chicago River); Columbus, OH (Scioto River); Dallas, TX (Trinity River); Denver, CO (South Platte River); Detroit, MI (Detroit River); El Paso, TX (Rio Grande); Fort Worth, TX (Trinity River); Indianapolis, IN (White River); Las Vegas, NV (Las Vegas Wash); Louisville, KY (Ohio River); Memphis, TN (Mississippi River); Nashville, TN (Cumberland River); Oklahoma City, OK (Oklahoma River); Phoenix, AZ (Salt River); and San Antonio, TX (San Antonio River). Together, these cities provide a nationally distributed and hydrologically diverse sample for examining how flood-induced network disruption and traffic dynamics shape urban accessibility.

\subsection*{Mobility demand and traffic simulation}

To characterize urban mobility patterns across the cities, we model traffic demand using an agent-based simulation approach (\cite{ambuhl2023understanding}). Traffic simulation captures how vehicles move through a road network under the influence of traveller behaviour, vehicle interactions, and network conditions, making it well suited for evaluating network performance under varying infrastructure and operational states (\cite{ambuhl2023understanding,muhlich2015use,dave2025local,pyatkova2019assessing}). In this study, we implement the traffic simulations in the Simulation of Urban MObility (SUMO) (\cite{lopez2018microscopic}), which represents vehicle movement in both space and time and generates outputs such as travel time and travel delay. Consistency across all cities is achieved through harmonized census-defined administrative units and standardized road-network data, which together define the simulation domain and network structure.

Census block groups (CBGs), the hierarchical administrative units defined by the United States Census Bureau (USCB), serve as the primary unit of analysis. To account for cross-boundary travel demand, we extend each city domain beyond its administrative boundary to include adjacent CBGs. CBG boundary and population data are obtained from (\cite{uscensus_tigerline_shapefiles}), which provides the geographic extent and demographic attributes of each unit. The road network is obtained from OpenStreetMap (OSM) (\cite{openstreetmap_website}), a routable geospatial database containing information on speed limits, number of lanes, traffic signals, and road class. Roads are categorized using standard OSM classes, including highway, primary, secondary, tertiary, and residential roads.

With the simulation domain and network structure defined, we next estimate city-level origin-destination commuting demand. Trip origins and destinations are derived from the Longitudinal Employer-Household Dynamics Origin-Destination Employment Statistics (LEHD–LODES) dataset maintained by the U.S. Census Bureau (\cite{uscensus_lodes}). LODES provides home-to-work flows between residential and workplace census blocks, which are aggregated to the CBG level to maintain consistency with the spatial resolution of the analysis. Because LODES reports commuting volumes at a daily rather than hourly scale, we temporally disaggregate these flows to construct hourly origin-destination demand for traffic simulation. The temporal profile for home-to-work trips is obtained from the American Community Survey (ACS) (\cite{acs_b08302}), which provides city-level journey-to-work patterns by transport mode at a semi-hourly resolution, whereas the temporal profile for work-to-home trips is obtained from the National Household Travel Survey (NHTS) \cite{nhts}. Daily LODES flows are then weighted by these directional temporal profiles to estimate hourly O-D trips between residential and workplace CBGs. Because the analysis focuses on road-network performance and accessibility, only car-based commute trips are considered, which contributes to $\sim$ 76\% of total commuting (\cite{census_commuting_acs_2025}).

Traffic simulations are conducted in SUMO using a mesoscopic multi-lane model (\cite{ambuhl2023understanding}). Roads are represented as discrete segments, and traffic propagates subject to link-level capacity and space constraints. Vehicle routes are assigned through a Dynamic User Equilibrium (DUE) formulation, which updates route choice in response to evolving congestion, travel time, and road capacity (\cite{ambuhl2023understanding,dave2025local}). Unlike static assignment, where routes remain fixed, DUE allows vehicles to adapt during the simulation and thereby captures traffic redistribution and queue formation more realistically. Intersections are modeled using a detailed right-of-way scheme. Routing follows a periodic stochastic shortest-path approach based on prevailing network conditions: 50\% of vehicles follow their initial route, whereas the remaining 50\% are re-routed every 6 minutes (\cite{ambuhl2023understanding}). This setup represents a mixed-routing environment in which some travellers remain on their original path while others adapt to changing traffic conditions. Additional details on model specification and validation are provided in ({Note S1}).

\subsection*{Accessibility quantification}

With the traffic simulations in place, we next quantify accessibility to critical facilities, focusing on healthcare as an essential urban service. Access to essential services is a key dimension of urban resilience, as transport disruption can directly limit the ability of residents to reach time-sensitive care (\cite{bruno2024universal,chen2025public,wu2025measuring}). Reliable healthcare access is especially important given the role of these facilities in both emergency response and routine healthcare, both of which become more vulnerable when external disruption impairs mobility (\cite{ceferino2024accessing,wassmer2025unveiling}). Healthcare POIsnts of interest (POIs) are extracted from the OpenStreetMap (OSM) geodatabase using its tag-based attribution system, where we consider hospitals as healthcare facilities (\cite{osm_planet_wiki}). Locations are identified with the query {`amenity': [`hospital'], `healthcare': [`hospital', `centre']}, and the resulting dataset is used to represent the number and spatial distribution of hospital facilities across the cities.

Accessibility measures are commonly grouped into two broad approaches: place-based measures, which evaluate accessibility from the perspective of locations, and individual-based measures, which focus on accessibility experienced by specific individuals or groups (\cite{vale2023accessibility,geurs2004accessibility}). In urban planning and infrastructure resilience, place-based accessibility is more commonly adopted because it enables consistent comparison across neighbourhoods, cities, and time periods. Such measures evaluate the ability of residents at a given location to reach relevant destinations through a transport system (\cite{fan2022equality,wu2025measuring,bruno2024universal,michelangeli2025inequality}). A place-based perspective is therefore appropriate for the present study, whose objective is to compare spatial variation in hospital access across census block groups and cities.

Within place-based accessibility, methods can be further distinguished by how opportunities are represented. Cumulative-opportunity measures quantify the number of destinations reachable within a specified travel threshold, but do not explicitly account for competition among users for those destinations (\cite{cheng2013measuring,pan2018evaluating,verma2025spatial,chen2025public}). Competition-based measures, by contrast, incorporate both the availability of reachable opportunities and the demand placed upon them, which is especially important for hospitals, where effective access declines as more people compete for finite service capacity (\cite{luo2009enhanced,del2022modeling,mao2013measuring}). Access is also not constant over time, as travel conditions and demand vary throughout the day (\cite{mao2025modeling,jarv2018dynamic}). To capture these temporal dynamics, we adopt a place-based, competition-sensitive measure of dynamic accessibility that accounts for both time-varying traffic conditions and population demand. We use a modified two-step floating catchment area (M2SFCA) metric to measure this dynamic, competition-based accessibility. Healthcare access is quantified using M2SFCA, as it evaluates reachability while accounting for competition for finite healthcare services under time-varying traffic conditions (\cite{delamater2013spatial,fang2025evaluating}). In this study, accessibility is weighted by a dynamic impedance term that reflects both travel-time cost and prevailing congestion along the route between origins and destinations.

To represent congestion in a spatially coherent manner in accessibility calculation, we partition the road network system into homogeneous traffic zones (HTZs). Such zoning is needed because congestion varies across the network, while edge-level traffic conditions are often too localized to provide a stable basis for impedance estimation (\cite{mahmassani2013urban,ramezani2015dynamics}). Therefore, we identify groups of adjacent road segments that exhibit similar temporal density patterns and treat them as contiguous zones representing coherent mesoscale traffic states (\cite{ramezani2015dynamics,yao2023novel}). The similarity between neighbouring edges is quantified by the correlation of hourly density time series, and the resulting adjacency-constrained graph is partitioned using the Leiden community detection algorithm (\cite{saeedmanesh2017dynamic,traag2019louvain}). The resulting zones are then refined by splitting disconnected components and merging small clusters with neighbouring zones based on the threshold of number of road edges, yielding the final HTZ configuration.

For each HTZ, we estimate a macroscopic fundamental diagram (MFD) using traffic observations at 5-min intervals within the zone to characterize the relationship between flow and density (\cite{ambuhl2023understanding}). The critical density is then identified as the density corresponding to maximum observed flow in the zone-specific MFD. Hourly congestion is subsequently represented by the relative density ratio $r_{z,t}$ (Eq.(\ref{Eq:1})).

\begin{equation}
r_{z,t}=\frac{k_{z,t}}{k_{\mathrm{crit},z}}
\label{Eq:1}
\end{equation}

where $r_{z,t}$ is the relative congestion ratio for HTZ $z$ at time $t$, $k_{z,t}$ is the observed density in zone $z$ at time $t$, and $k_{\mathrm{crit},z}$ is the critical density derived from the MFD. This ratio provides a normalized measure of the congestion state for use in the accessibility impedance function.

For each origin--destination pair, the congestion term is represented by the mean relative density ratio across the HTZs traversed by the shortest path at time \(t\) as shown in (Eq.(\ref{Eq:2})).

\begin{equation}
\bar{r}_{ij,t}=\frac{1}{\left|Z_{ij,t}\right|}\sum_{z \in Z_{ij,t}} r_{z,t}
\label{Eq:2}
\end{equation}

where $Z_{ij,t}$ denotes the set of HTZs crossed by the route between origin CBG (all road intersection in that CBG) $i$ and hospital $j$ at time $t$. 

The impedance function is then defined as a joint function of simulated travel time and route-level congestion state (Eq.(\ref{Eq:3})),

\begin{equation}
f_{ij,t}=\frac{\exp\left[-\beta\left(\bar{r}_{ij,t}\right)\right]}{T_{ij,t}}
\label{Eq:3}
\end{equation}

where, $T_{ij,t}$ is the simulated travel time in minutes between origin CBG $i$ and hospital $j$ at time $t$, and $\beta$ is the decay parameter (0.1). Under this formulation, accessibility declines with both increasing travel time and increasingly congested traffic conditions.

Using this impedance term, the first step of the M2SFCA calculation computes a time-dependent supply-to-demand ratio for each hospital (Eq.(\ref{Eq:4})),

\begin{equation}
R_{j,t}=\frac{S_j}{\sum_{i \in \omega_j} P_i f_{ij,t}}
\label{Eq:4}
\end{equation}

where $R_{j,t}$ is the supply-to-demand ratio of hospital $j$ at the time $t$, $S_j$ denotes hospital supply, $P_i$ is the population of origin CBG $i$ and $\omega_j$ is the set of origin CBGs within the hospital catchment (reachable inside 30 min threshold) . In the second step, accessibility at each origin is calculated by summing the weighted hospital-specific ratios of all reachable hospitals (Eq.(\ref{Eq:5})),

\begin{equation}
A_{i,t}=\sum_{j \in J_i} R_{j,t} f_{ij,t}
\label{Eq:5}
\end{equation}

where $A_{i,t}$ is the accessibility of origin CBG $i$ at the time $t$, and $J_i$ is the set of hospitals reachable from origin CBG $i$ within the defined travel-time threshold, in our case we considered 30 min as threshold. This procedure yields an hourly, competition-sensitive measure of healthcare accessibility that incorporates both demand pressure and time-varying network congestion. The city level aggregate accessibility is quantified as (Eq.(\ref{Eq:6})):

\begin{equation}
\bar{A}_{t}=\frac{1}{N}\sum_{i=1}^{N} A_{i,t}
\label{Eq:6}
\end{equation}

where $\bar{A}_{t}$ is the mean accessibility of the city at time t. $A_{i,t}$ is the accessibility of CBG $i$ at the time $t$ and $N$ is the total number of CBGs in the city.

To quantify congestion-related accessibility loss, we compare accessibility during peak traffic hours with the corresponding accessibility under non-peak conditions. City-level hourly accessibility is obtained by averaging CBG-level accessibility across all CBGs, and relative loss is then calculated as the percentage reduction from non-peak to peak-hour accessibility (Eq.(\ref{Eq:7})).

\begin{equation}
\%\mathrm{Loss}_{congestion}=
\frac{\bar{A}_{t_{\mathrm{off-peak}}}-\bar{A}_{t_{\mathrm{peak}}}}
{\bar{A}_{t_{\mathrm{off-peak}}}}
\times 100
\label{Eq:7}
\end{equation}

\subsection*{Flood hazard characterization}
Flooding is a major hazard for road transportation systems, causing both direct and indirect disruption to network performance and, consequently, accessibility to critical facilities (\cite{dave2021extreme,wassmer2025unveiling,lin2024assessing,dave2025local}). In this study, flood hazard is classified by its dominant generating mechanism into pluvial, fluvial, and coastal flooding. Pluvial flooding is driven by extreme precipitation interacting with local topography and surface conditions, fluvial flooding by riverine overflow beyond channel banks under high discharge, and coastal flooding by elevated tidal levels and storm surges (\cite{wing202430,dave2025local}).

Flood hazard characterization is based on the 100-year return-period flood data developed by Fathom (\cite{wing202430}). The 100-year return period is used to provide a consistent hazard benchmark for comparing flood-induced disruption across cities and flood mechanisms. These data provide flood depth and extent estimates for pluvial, fluvial, and coastal processes at 30 m spatial resolution, depending on the mechanisms relevant to each city. The dataset is generated using the LISFLOOD-FP 1D–2D hydrodynamic model, which simulates water movement across river channels, floodplains, and urban surfaces (\cite{wing202430}). We use the 2020 baseline conditions in all analyses.
For inland cities, pluvial and fluvial flood layers are extracted within the defined city domain, whereas for coastal cities pluvial, fluvial, and coastal layers are all included. We additionally construct a composite flood scenario by taking the union of the relevant hazard layers and retaining the maximum flood depth in overlapping areas (\cite{del2025composite}). In the absence of explicitly modeled non-linear interactions among flood-generating mechanisms, this composite scenario serves as a practical representation of compound flooding.

\subsection*{Flood-induced accessibility disruption}
Flood-induced disruption to road network is quantified by integrating flood hazard data with the simulated road network. Flood extent and depth layers are overlaid on the network to identify exposed road segments and assign inundation depth to each affected link. Road segments are treated as closed when flood depth exceeds 0.3 m (\cite{martinez2017new,dave2025local,pyatkova2019assessing}), whereas shallower flooding is represented through a depth–disruption relationship that reduces traffic performance (\cite{pyatkova2019assessing,pregnolato2017impact}). In doing so, the analysis captures both road closure and flood-related slowdown, reflecting the two main ways in which inundation perturbs urban traffic conditions.

Flood-induced disruptions are incorporated into the mesoscopic traffic simulation by modifying network conditions on affected road segments. Roads exceeding the closure threshold are blocked by placing rerouting rules upstream of the inundated segment, allowing vehicles to divert before reaching an impassable link (\cite{pyatkova2019assessing}). Roads exposed to shallower flooding remain open, but operate with reduced speed and capacity according to the depth–disruption relationship. Under these modified network conditions, travel demand is kept the same as in the baseline scenario, assuming unchanged trip-making behaviour during the simulated flood event. The model is then re-simulated for pluvial, fluvial, coastal, and composite flood scenarios. To examine system response under already stressed conditions, flood-related closures are imposed during the morning peak (07:00–10:00) and evening peak (17:00–20:00). Travel times and other traffic performance metrics from these simulations are subsequently used to quantify healthcare accessibility under flood exposure. 

Accessibility under flood conditions is quantified using the same M2SFCA framework (\cite{delamater2013spatial,fang2025evaluating}), while accounting for flood-induced road closure, reduced speed under shallow inundation, and direct disruption to healthcare access. We also account for hospitals that become inundated or unreachable because no alternative route remains available. Using these modified network conditions, dynamic accessibility is quantified for each flood scenario (Eq.\ref{Eq:8}). To evaluate the effect of flooding on accessibility, we compare the time-varying accessibility profiles under flooded and baseline conditions (without flood). Accessibility loss is then quantified by comparing the temporal accessibility profiles under flooded and baseline conditions. For each city and scenario, the area under the hourly accessibility curve is calculated for the morning peak, evening peak, and full day using the trapezoidal rule (Eq.\ref{Eq:9}). Loss is defined as the difference between the baseline and flooded areas under the curve, and relative loss is expressed as the percentage reduction with respect to baseline accessibility (Eq.\ref{Eq:10}). This approach captures not only the magnitude of accessibility decline, but also its persistence over time.

\begin{equation}
\bar{A}_{s}(t)=\frac{1}{N}\sum_{i=1}^{N} A_{i,s}(t)
\label{Eq:8}
\end{equation}

where $\bar{A}_{s}(t)$ is the mean accessibility at hour $t$ under scenario $s$ (pluvial, fluvial, coastal and composite), $A_{i,s}(t)$ is the accessibility of CBG $i$ at hour $t$ under scenario $s$, and $N$ is the total number of CBGs in the city. 

\begin{equation}
\bar{A}_{s}^{[t_1,t_2]}=\int_{t_1}^{t_2} \bar{A}_{s}(t)\,dt
\label{Eq:9}
\end{equation}

where $\bar{A}_{s}^{[t_1,t_2]}$ is the accessibility aggregated over the selected time period, such as the morning peak, evening peak, or full day.

\begin{equation}
\% \mathrm{Loss}^{[t_1,t_2]}_{s}=
\frac{\bar{A}_{\mathrm{base}}^{[t_1,t_2]}-\bar{A}_{s}^{[t_1,t_2]}}
{\bar{A}_{\mathrm{base}}^{[t_1,t_2]}}
\times 100
\label{Eq:10}
\end{equation}

\begin{equation}
\% \mathrm{Loss (A)}_{flood}=
\frac{\bar{A}_{\mathrm{base}}-\bar{A}_{flood}}
{\bar{A}_{\mathrm{base}}}
\times 100
\label{Eq:10}
\end{equation}

where $\bar{A}_{\mathrm{base}}$ is the baseline accessibility aggregated over the selected time period such as the morning peak, evening peak, or full day and $\bar{A}_{s}^{[t_1,t_2]}$ is the accessibility under flood scenario $s$. Finally, for analysis we considered the maximum value of accessibility loss among the selected period.

\subsection*{Drivers of accessibility loss and vulnerability typology}

To explain why flood-induced accessibility loss differs across cities and scenarios, we decompose the observed loss into four mechanistic drivers based on structural and functional vulnerability of city when urban road-healthcare system exposed to flooding: road-network exposure (E),fragility of hospital access (F), congestion amplification (C), and hospital isolation (H). These variables capture complementary dimensions of vulnerability, spanning direct hazard exposure, network-level sensitivity to hospital access, traffic-system response, and destination-level disruption. Together, these variables describe how direct hazard exposure propagates through the transport system to produce unequal accessibility outcomes.

The first driver, road-network exposure $(E)$, measures the structural imprint of flooding on the transport network. Rather than quantifying exposure only by the number or length of affected roads, $E$ weights flood-affected links by edge betweenness centrality to reflect their structural importance for network connectivity (\cite{girvan2002community,arrighi2019preparedness}). Edge betweenness is computed on the baseline road network using free-flow travel time as the edge weight, so that $E$ captures how flooding affects the inherent importance of network links rather than traffic conditions that arise from congestion (Eq.\ref{Eq:11}). Flood-closed roads contribute their full betweenness weight, whereas roads operating under reduced speed contribute a fraction of their betweenness proportional to the reduction in allowable speed. 

\begin{equation}
E=\frac{\sum_{e \in \mathcal{F}_{\mathrm{closed}}} b_e+\sum_{e \in \mathcal{F}_{\mathrm{reduced}}} \alpha_e b_e}{\sum_{e \in \mathcal{N}} b_e}
\label{Eq:11}
\end{equation}
 
where $b_e$ denotes the betweenness centrality of edge $e$, and $\alpha_e$ is the fractional speed reduction under flooding, as shown in (Eq.\ref{Eq:12}).
\begin{equation}
\alpha_e = 1-\frac{v^{\mathrm{reduced}}_e}{v^{\mathrm{free}}_e}
\label{Eq:12}
\end{equation}

so the exposure $(E)$ captures the extent to which flooding disrupts functionally important roads rather than peripheral links alone.

The second driver, network fragility with respect to hospital access $(F)$, measures the extent to which flood-induced road disruption increases travel time to the nearest hospital. Unlike exposure $(E)$, which captures the structural imprint of flooding on the network itself, $F$ reflects the sensitivity of access to healthcare under disrupted routing conditions (\cite{gangwal2023critical,dong2022modest}). For each zone (CBGs), we compute the shortest travel time to the nearest hospital under free-flow and flood conditions, and express fragility as the ratio of flooded to free-flow travel time. These zone-level ratios are then aggregated using population weights of CBG as shown in (Eq.\ref{Eq:13}).

\begin{equation}
F=\frac{\sum_{i=1}^{N} P_i \left(\dfrac{T^{\mathrm{flood}}_{i}}{T^{\mathrm{free-flow}}_{i}}\right)}
{\sum_{i=1}^{N} P_i}
\label{Eq:13}
\end{equation}
  
where $T^{\mathrm{free-flow}}_{i}$ and $T^{\mathrm{flood}}_{i}$ denote the nearest-hospital travel times from CBG $i$ under free-flow and flood conditions, respectively, and $P_i$ denotes the population of CBG $i$. Higher values of $F$ indicate a more fragile network, in which flooding produces larger increases in hospital travel time for the exposed population.

The third driver, congestion amplification $(C)$, measures the extent to which flooding increases congestion burden beyond baseline conditions. Whereas exposure $(E)$ captures the structural imprint of flooding on the network and fragility $(F)$ captures the resulting sensitivity of hospital reachability, $C$ reflects the additional traffic delay generated by flood disruption during already congested periods which reflects the functional vulnerability of the road network system. For each city, congestion burden is defined as the mean time loss per inserted vehicle during the morning peak, computed as the ratio of total time loss to the number of inserted vehicles on the road edge (\cite{lin2024assessing,sumoEdgeDataDoc}). Congestion amplification is then expressed as the absolute difference between flooded and baseline congestion burden (Eq.\ref{Eq:14}).

\begin{equation}
C=\Delta c=c^{\mathrm{flood}}-c^{\mathrm{base}}
\label{Eq:14}
\end{equation}

where, $c^\mathrm{base}$ and $c^\mathrm{flood}$denote the per-vehicle congestion burden under baseline and flood conditions, respectively, with $c$ quantified as, 

\begin{equation}
c=\frac{\mathrm{TL}}{\mathrm{INS}}
\label{Eq:15}
\end{equation}

where $TL$ is the total time loss accumulated over the analysis period and $INS$ is the number of inserted vehicles. Higher values of $C$ indicate stronger amplification of congestion under flooding.

The fourth driver, hospital isolation $(H)$, measures the extent to which flooding reduces the population catchment of hospitals. Unlike fragility $(F)$, which captures flood-induced changes in travel time to the nearest hospital from the population perspective and thus reflects the functional sensitivity of hospital access within a city, $H$ captures the loss of hospital reach from the facility perspective (\cite{wassmer2025unveiling}). For each hospital $h$, the catchment population within the specified travel-time threshold is computed under both free-flow and flood conditions. The hospital-specific catchment ratio is then defined as (Eq.\ref{Eq:16}).

\begin{equation}
r_h=\frac{\omega_h^{\mathrm{flood}}}{\omega_h^{\mathrm{free-flow}}}
\label{Eq:16}
\end{equation}

where $\omega_h^{\mathrm{free-flow}}$ and $\omega_h^{\mathrm{flood}}$ denote the free-flow  and flooded condition catchment populations of hospital $h$, respectively. Hospital isolation is subsequently calculated as (Eq.\ref{Eq:17}).

\begin{equation}
H=1-\frac{1}{M}\sum_{h=1}^{M} r_h
\label{Eq:17}
\end{equation}
where $M$ is the number of hospitals with non-zero free-flow condition catchment. Higher values of $H$ indicate stronger reduction in the population that can reach hospitals within the defined travel-time threshold (30 min).

To enable systematic comparison and interpretation of accessibility loss across cities, the four mechanistic drivers are further translated into a vulnerability typology based on whether each driver lies above or below the cross-city median. This median-based classification provides a concise and interpretable summary of the dominant processes through which flooding reduces hospital accessibility, while retaining the distinction between road system exposure, access fragility, congestion amplification and hospital isolation. The grouping further helps identify whether observed accessibility loss is primarily driven by structural vulnerability or functional vulnerability of city or by compound effects across multiple drivers. Cities are assigned to one of four merged vulnerability classes. Compound critical cities have all four drivers above the median, indicating simultaneous structural disruption (High E and F), functional vulnerability (High C and H). Functional vulnerability cities have both congestion amplification and hospital isolation above the median, but not both structural drivers simultaneously. Structural vulnerability characterizes cities where exposure and fragility dominate in the absence of strong amplification effects of functional vulnerability drivers. Resilient cities have all four drivers at or below the median, indicating relatively limited accessibility disruption despite flooding. The vulnerability typology is further supported by statistical testing ( Kruskal–Wallis (H statistic), Eta-squared ($\eta {2}$) effect size, Spearman rank correlation ($\rho$)), which shows that accessibility loss differs significantly across mechanism classes and increases systematically with mechanism stacking, details are provided in ({Note S2}). 

\subsection*{Inequality measures}
To assess the broader distributional consequences of accessibility loss, we quantify inequality in hospital accessibility across census block groups (CBGs) using the Gini coefficient (\cite{bruno2024universal,sanders2023large}). In this context, the Gini coefficient captures the extent to which access to hospitals is unevenly distributed within each city. To account for variation in CBG population, we use a population-weighted Gini coefficient so that the resulting measure reflects inequality across residents rather than across spatial units alone. After ordering CBGs by accessibility in ascending order, the cumulative population share $W_{k}$ and cumulative accessibility share $L_{k}$ are defined as (Eq.\ref{Eq:18}).
\begin{equation}
W_k=\frac{\sum_{i=1}^{q} p_i}{\sum_{i=1}^{N} p_i},
\qquad
L_k=\frac{\sum_{i=1}^{q} p_i A_i}{\sum_{i=1}^{N} p_i A_i}
\label{Eq:18}
\end{equation}

where $A_i$ denotes hospital accessibility in CBG $i$, $p_i$ is the population of CBG $i$, and $N$ is the number of CBG in the city. The population-weighted Gini coefficient (ranging from 0 to 1) is then calculated from the weighted Lorenz curve as (Eq.\ref{Eq:19}).

\begin{equation}
G = 1 - 2 \sum_{q=1}^{N} \frac{L_k + L_{k-1}}{2}\,(W_k - W_{k-1})
\label{Eq:19}
\end{equation}

with $W_0 = L_0 = 0$. Higher values of $G$ indicate greater inequality in the distribution of hospital accessibility across residents.

To further examine the social distribution of hospital accessibility, we quantify the Concentration Index (CI) at the census block group (CBG) level. Unlike the Gini coefficient, which measures overall inequality, the concentration index captures whether accessibility is systematically concentrated among more or less socially vulnerable communities.The Social Vulnerability Index (SVI) data is collected from (\cite{flanagan2011social}) at CBG scale. CBGs are ranked from lower to higher social vulnerability, and the concentration index is computed in population-weighted form so that the measure reflects how accessibility is distributed across residents along the vulnerability gradient \cite{odonnell2008_concentration_index,larrabee2022racialized}. The fractional population rank of CBG $i$ is defined as (Eq.\ref{Eq:20}).

\begin{equation}
\phi_i=\frac{1}{P}\left(\sum_{j=1}^{i-1} p_j+\frac{1}{2}p_i\right)
\label{Eq:20}
\end{equation}

where $\phi_i$ is the fractional population rank of CBG $i$ in the cumulative distribution of social vulnerability, $CD_i$ is the population of CBG $i$, and ${P}=\sum_{i=1}^{N} CD_i$
is the total city population. The population-weighted mean accessibility is given by (Eq.\ref{Eq:21}).
 
\begin{equation}
\mu=\frac{1}{P}\sum_{i=1}^{N} CD_i A_i
\label{Eq:21}
\end{equation}

where $A_i$ is hospital accessibility in CBG $i$, and $\mu$ is the population-weighted mean accessibility.

The concentration index is calculated as (Eq.\ref{Eq:22}).
\begin{equation}
CI=\frac{2}{\mu P}\sum_{i=1}^{N} CD_i A_i \phi_i - 1
\label{Eq:22}
\end{equation}

Negative values of $CI$ indicate that hospital accessibility is concentrated among more socially vulnerable groups, whereas positive values indicate concentration among less vulnerable groups. Values close to zero indicate little systematic concentration along the vulnerability gradient.

\section*{Data availability}
The datasets used in this study were obtained from publicly accessible repositories and institutional or municipal data portals. Road network data were derived from OpenStreetMap (OSM) (\cite{osm_planet_wiki}). City and census block group boundary data were obtained from the U.S. Census (\cite{uscensus}). Traffic demand data were obtained from the Longitudinal Employer-Household Dynamics Origin-Destination Employment Statistics (LEHD-LODES) dataset (\cite{uscensus_lodes}). Hourly trip pattern information was compiled from the American Community Survey (ACS) and the National Household Travel Survey (NHTS) (\cite{acs_b08302}). Flood hazard data, including fluvial, pluvial, and coastal flood extents, were obtained from Fathom Global (\cite{wing202430}). Social vulnerability data were obtained from the CDC/ATSDR Social Vulnerability Index (SVI) (\cite{flanagan2011social}). Any processed datasets generated during the analysis can be made available by the authors upon reasonable request, subject to the access and licensing terms of the original data providers.

\section*{Code availability}

Python scripts used for the analysis and figure generation are publicly available on \url{https://github.com/raviraj-dave96/Dynamical-healthcare-accessibility-losses-under-compounding-hazards}.

\section*{Main Figures}
\clearpage

\clearpage

\textbf{Figure 1: Spatial coverage and traffic patterns of study cities across coastal and inland settings with conceptual workflow.} \textbf{(a)} Locations of the 30 metropolitan study areas across the contiguous United States, with marker size indicating city population. \textbf{(b)} The maps depict the simulated peak-hour traffic patterns for representative coastal cities (Boston, Baltimore, Chicago and Los Angeles) and inland cities (Houston, Dallas, Memphis and Indianapolis). Road segments are colored by time loss, from low (white) to high (red), showing contrasting congestion structures across urban networks. \textbf{(c)} Conceptual framework, including urban mobility simulation, dynamic accessibility estimation to critical facilities at the census block group (CBG) level, assessment of flood exposure (pluvial, fluvial, coastal and composite) on the road network and critical facilities, estimation of flood-induced accessibility loss, and evaluation of the resulting spatial and social inequality in accessibility.

\clearpage

\clearpage

\textbf{Figure 2: Traffic congestion reduces access to healthcare and amplifies intra-urban inequality.} \textbf{(a)} Daily profiles of population-weighted healthcare accessibility for four representative cities (Boston, Los Angeles, Chicago, and Houston), highlighting the percent reduction from non-peak hour to peak hour. \textbf{(b)} City level severity of congestion induced accessibility loss (from non-peak to peak hours). \textbf{(c)} Spatial variation of non-peak-hour accessibility with in city, expressed as z-scores to highlight neighborhoods with above- and below-average access. \textbf{(d)} Census block group (CBG)–level spatial variation in percent change accessibility from non-peak to peak hours, indicating localized losses and gains. $N$ represents the number of CBGs. \textbf{(e)} Distributions of within city accessibility during peak and non-peak hours for major coastal and inland cities.

%

\clearpage

\clearpage

\textbf{Figure 3: Disparate flood drivers produce distinct road-network disruption.} Blue bars show total inundation volume within each city, and orange bars show the share of road length inundated (\% roads flooded); cities in each panel are ordered by inundation volume. Insets illustrate, for the most exposed representative city in each hazard type, the spatial distribution of flood depth and the corresponding flooded road segments over the traffic network. \textbf{(a)} Pluvial flooding, shown for Houston; \textbf{(b)} fluvial flooding, shown for Portland; and \textbf{(c)} coastal flooding, shown for Boston, the cities are selected as their road disruptions are highest in each flood.

\clearpage

\clearpage

\textbf{Figure 4: Flooding intensifies healthcare accessibility loss and spatial inequality across cities and flood mechanisms.} \textbf{(a)} Spatial variation in hospital accessibility under pluvial, fluvial, coastal and composite flooding for Boston and Houston, shown as percentage change relative to the baseline no-flood condition. Red denotes accessibility loss and blue denotes accessibility gain; hospital locations are overlaid and city boundaries are outlined. \textbf{(b)} Distributions of census block group level healthcare accessibility under baseline and flood conditions, showing downward shifts and broader dispersion under more severe flood scenarios. \textbf{(c)} Temporal dynamics of citywide accessibility in Boston and Houston under each flood mechanism, showing the strongest reductions during peak travel periods and the highest losses under composite flooding. \textbf{(d)} City-level severity of flood-induced accessibility loss across flood mechanisms, highlighting substantial inter-city variation. \textbf{(e)} Shifts in city-level severity of accessibility loss from congestion to flooding for each hazard type. Ribbons connect each city’s congestion-loss severity to its flood-loss severity and are colored by coastal versus inland classification.

\clearpage

\clearpage

\textbf{Figure 5: Multi-driver vulnerability typologies explain variation in accessibility loss across cities and flood scenarios.} \textbf{(a)} Inter-city comparison of the four mechanistic drivers of accessibility loss, hospital isolation (H), congestion amplification (C), fragility to hospital access (F) and exposure to road (E), for pluvial, fluvial, coastal and composite flooding. Color intensity indicates the level of each driver within each city and hazard scenario. \textbf{(b)} Vulnerability profiles of cities across flood scenarios. Points are positioned by exposure to road (E) and fragility to hospital access (F), colored by accessibility loss (\%), scaled by congestion amplification (C), and symbolized by inferred vulnerability typology. Reference lines denote median thresholds used to distinguish low and high driver conditions. \textbf{(c)} Accessibility-loss severity across vulnerability typologies for each flood scenario. Box plots show increasing losses from resilient to compound-vulnerability classes, indicating that cities with multiple elevated drivers experience the greatest accessibility disruption.

\clearpage

\clearpage

\textbf{Figure 6: Flooding amplifies spatial and social inequality in hospital accessibility} Gini index (GI) and concentration index (CI) of accessibility loss are shown for \textbf{(a)} pluvial, \textbf{(b)} fluvial, \textbf{(c)} coastal and \textbf{(d)} composite flooding. The left dumbbell plot in each panel compares population-weighted spatial inequality in accessibility under baseline conditions (small grey dots) and flood conditions (large purple dots), with adjacent annotations indicating the change in GI from baseline to the corresponding flood scenario with color grading representing the ($\%Loss(A)_{flood}$). The right dumbbell plot shows the corresponding social inequality in accessibility loss using the concentration index, with small grey dots denoting baseline values and colored dots denoting flood conditions; adjacent annotations report the change in CI from baseline. Negative CI values indicate greater accessibility loss for socially vulnerable groups, whereas positive values indicate greater loss for less vulnerable groups.

\section*{Abbreviation}
    
City abbreviations used throughout the manuscript are as follows: Austin (AU), Baltimore (BA), Boston (BO), Charlotte (CL), Chicago (CH), Columbus (CO), Dallas (DA), Denver (DE), Detroit (DT), El Paso (EP), Fort Worth (FW), Houston (HO), Indianapolis (IN), Jacksonville (JA), Las Vegas (LV), Los Angeles (LA), Louisville (LO), Memphis (ME), Nashville (NA), New York (NY), Oklahoma City (OK), Philadelphia (PH), Phoenix (PX), Portland (PO), San Antonio (SA), San Diego (SD), San Jose (SJ), San Francisco (SF), Seattle (SE), and Washington, DC (WD).


\section*{Acknowledgments}
This research was supported by the U.S. Department of Defense Strategic Environmental Research and Development Program (SERDP) under award RC20-1183, with additional support from Northeastern University. The authors thank the current members of Northeastern University’s Sustainability and Data Sciences Laboratory (SDS Lab) and AI for Climate and Sustainability (AI4CaS) for helpful discussions and feedback. The authors also gratefully acknowledge the Fathom Global for providing flood hazard data used in this study.

\section*{Competing interests}
Authors declare no competing interests.
\section*{Author contributions}

Conceptualization:  R.D., D.M.T., D.Z., A.R.G; Design of Experiments: R.D., D.M.T., D.Z.; Analysis:  R.D., D.M.T., D.Z.; Writing - Original Draft: R.D., D.M.T., D.Z.; Writing -
Review \& Editing: R.D., D.M.T., D.Z., U.B., A.R.G

\bibliographystyle{unsrtnat}
\bibliography{references}  

@article{vale2023accessibility,
  title={Accessibility inequality across Europe: a comparison of 15-minute pedestrian accessibility in cities with 100,000 or more inhabitants. NPJ Urban Sustain. 2023; 3 (1): 1--13},
  author={Vale, D and Lopes, A Soares},
  journal={DOI},
  volume={10},
  pages={55},
  year={2023}
}

@article{geurs2004accessibility,
  title={Accessibility evaluation of land-use and transport strategies: review and research directions},
  author={Geurs, Karst T and Van Wee, Bert},
  journal={Journal of Transport geography},
  volume={12},
  number={2},
  pages={127--140},
  year={2004},
  publisher={Elsevier}
}

@article{chen2025public,
  title={Public transport accessibility in villages in and around major Chinese cities},
  author={Chen, Zihua and Li, Xiaowei and Liu, Bingzhi and Wang, Shaohua and Li, Xiao and Li, Jiaxin and Yang, Xiaohui and Wang, Zhenbo},
  journal={Nature Cities},
  volume={2},
  number={8},
  pages={749--758},
  year={2025},
  publisher={Nature Publishing Group US New York}
}

@article{ceferino2024accessing,
  title={Accessing Acute Care Hospitals in the San Francisco Bay after a Major Hayward Earthquake},
  author={Ceferino, Luis and Kukunoor, Charan and Mao, Dan and Xu, Xinlu and Wu, Jingzhe and Zsarn{\'o}czay, Adam},
  journal={Preprint at https://doi. org/10.31224/3605},
  year={2024}
}

@article{wassmer2025unveiling,
  title={Unveiling hidden risks in healthcare from flood-induced transportation disruption in Germany},
  author={Wassmer, Jonas and Bryant, Seth and Schimansky, Paul and Keegan, Lindsay T and Pregnolato, Maria and Kurths, J{\"u}rgen and Marwan, Norbert and Merz, Bruno},
  journal={Communications Earth \& Environment},
  volume={6},
  number={1},
  pages={676},
  year={2025},
  publisher={Nature Publishing Group UK London}
}

@misc{osm_planet_wiki,
  title        = {Planet.osm},
  author       = {{OpenStreetMap Wiki contributors}},
  howpublished = {\url{https://wiki.openstreetmap.org/wiki/Planet.osm}},
  note         = {Accessed 2026-03-13},
  year         = {2026}
}

@article{bruno2024universal,
  title={A universal framework for inclusive 15-minute cities},
  author={Bruno, Matteo and Monteiro Melo, Hygor Piaget and Campanelli, Bruno and Loreto, Vittorio},
  journal={Nature Cities},
  volume={1},
  number={10},
  pages={633--641},
  year={2024},
  publisher={Nature Publishing Group US New York}
}

@article{wu2025measuring,
  title={Measuring global human accessibility to essential daily necessities and services},
  author={Wu, Shengbiao and Chen, Bin and An, Jiafu and Nelson, Andrew and Dai, Fan and Lin, Chen and Gong, Peng},
  journal={Nature communications},
  volume={16},
  number={1},
  pages={10709},
  year={2025},
  publisher={Nature Publishing Group UK London}
}

@article{fan2022equality,
  title={Equality of access and resilience in urban population-facility networks},
  author={Fan, Chao and Jiang, Xiangqi and Lee, Ronald and Mostafavi, Ali},
  journal={npj Urban Sustainability},
  volume={2},
  number={1},
  pages={9},
  year={2022},
  publisher={Nature Publishing Group UK London}
}

@article{michelangeli2025inequality,
  title={Inequality in access to urban amenities},
  author={Michelangeli, Alessandra and {\"O}sth, John and Toger, Marina and T{\"u}rk, Umut},
  journal={npj Urban Sustainability},
  volume={5},
  number={1},
  pages={54},
  year={2025},
  publisher={Nature Publishing Group UK London}
}

@article{cheng2013measuring,
  title={Measuring urban job accessibility with distance decay, competition and diversity},
  author={Cheng, Jianquan and Bertolini, Luca},
  journal={Journal of Transport geography},
  volume={30},
  pages={100--109},
  year={2013},
  publisher={Elsevier}
}

@article{pan2018evaluating,
  title={Evaluating the accessibility of healthcare facilities using an integrated catchment area approach},
  author={Pan, Xiaofang and Kwan, Mei-Po and Yang, Lin and Zhou, Shunping and Zuo, Zejun and Wan, Bo},
  journal={International Journal of Environmental Research and Public Health},
  volume={15},
  number={9},
  pages={2051},
  year={2018},
  publisher={MDPI}
}

@article{verma2025spatial,
  title={Spatial Access of America: Multiple indicators of accessibility to opportunities},
  author={Verma, Rajat and Mittal, Shagun and Ukkusuri, Satish V},
  journal={Scientific Data},
  volume={12},
  number={1},
  pages={1223},
  year={2025},
  publisher={Nature Publishing Group UK London}
}

@article{delamater2013spatial,
  title={Spatial accessibility in suboptimally configured health care systems: A modified two-step floating catchment area (M2SFCA) metric},
  author={Delamater, Paul L},
  journal={Health \& place},
  volume={24},
  pages={30--43},
  year={2013},
  publisher={Elsevier}
}

@article{luo2009enhanced,
  title={An enhanced two-step floating catchment area (E2SFCA) method for measuring spatial accessibility to primary care physicians},
  author={Luo, Wei and Qi, Yi},
  journal={Health \& place},
  volume={15},
  number={4},
  pages={1100--1107},
  year={2009},
  publisher={Elsevier}
}

@article{mao2013measuring,
  title={Measuring spatial accessibility to healthcare for populations with multiple transportation modes},
  author={Mao, Liang and Nekorchuk, Dawn},
  journal={Health \& place},
  volume={24},
  pages={115--122},
  year={2013},
  publisher={Elsevier}
}

@article{mao2025modeling,
  title={Modeling time-varying spatial accessibility to healthcare: A system dynamic approach},
  author={Mao, Liang},
  journal={Health \& Place},
  volume={91},
  pages={103416},
  year={2025},
  publisher={Elsevier}
}

@article{jarv2018dynamic,
  title={Dynamic cities: Location-based accessibility modelling as a function of time},
  author={J{\"a}rv, Olle and Tenkanen, Henrikki and Salonen, Maria and Ahas, Rein and Toivonen, Tuuli},
  journal={Applied geography},
  volume={95},
  pages={101--110},
  year={2018},
  publisher={Elsevier}
}

@article{fang2025evaluating,
  title={Evaluating Dynamic Accessibility of Transportation Network Under Extreme Rainfall and Flooding: An Integrated Framework},
  author={Fang, Xinyi and Lu, Linjun and Hong, Yilin and Sun, Poly ZH},
  journal={IEEE Internet of Things Journal},
  year={2025},
  publisher={IEEE}
}

@article{mahmassani2013urban,
  title={Urban network gridlock: Theory, characteristics, and dynamics},
  author={Mahmassani, Hani S and Saberi, Meead and Zockaie, Ali},
  journal={Procedia-Social and Behavioral Sciences},
  volume={80},
  pages={79--98},
  year={2013},
  publisher={Elsevier}
}

@article{ramezani2015dynamics,
  title={Dynamics of heterogeneity in urban networks: aggregated traffic modeling and hierarchical control},
  author={Ramezani, Mohsen and Haddad, Jack and Geroliminis, Nikolas},
  journal={Transportation Research Part B: Methodological},
  volume={74},
  pages={1--19},
  year={2015},
  publisher={Elsevier}
}

@article{yao2023novel,
  title={A novel self-adaption macroscopic fundamental diagram considering network heterogeneity},
  author={Yao, Wenbin and Chen, Nuo and Su, Hongyang and Hu, Youwei and Jin, Sheng and Rong, Donglei},
  journal={Physica A: Statistical Mechanics and its Applications},
  volume={613},
  pages={128531},
  year={2023},
  publisher={Elsevier}
}

@article{saeedmanesh2017dynamic,
  title={Dynamic clustering and propagation of congestion in heterogeneously congested urban traffic networks},
  author={Saeedmanesh, Mohammadreza and Geroliminis, Nikolas},
  journal={Transportation research procedia},
  volume={23},
  pages={962--979},
  year={2017},
  publisher={Elsevier}
}

@article{traag2019louvain,
  title={From Louvain to Leiden: guaranteeing well-connected communities},
  author={Traag, Vincent A and Waltman, Ludo and Van Eck, Nees Jan},
  journal={Scientific reports},
  volume={9},
  number={1},
  pages={5233},
  year={2019},
  publisher={Nature Publishing Group UK London}
}

@misc{uscensus,
  author       = {{United States Census Bureau}},
  title        = {United States Census Bureau},
  year         = {2026},
  url          = {https://www.census.gov/},
  note         = {Accessed: 2026-03-17}
}

@article{ambuhl2023understanding,
  title={Understanding congestion propagation by combining percolation theory with the macroscopic fundamental diagram},
  author={Amb{\"u}hl, Lukas and Menendez, Monica and Gonz{\'a}lez, Marta C},
  journal={Communications Physics},
  volume={6},
  number={1},
  pages={26},
  year={2023},
  publisher={Nature Publishing Group UK London}
}

@article{muhlich2015use,
  title={Use of microsimulation for examination of macroscopic fundamental diagram hysteresis patterns for hierarchical urban street networks},
  author={M{\"u}hlich, Nicolas and Gayah, Vikash V and Menendez, Monica},
  journal={Transportation Research Record},
  volume={2491},
  number={1},
  pages={117--126},
  year={2015},
  publisher={SAGE Publications Sage CA: Los Angeles, CA}
}

@article{dave2025local,
  title={Local damages drive network-wide costs in compound flood-prone coastal city},
  author={Dave, Raviraj and Sen, Sushobhan and Maji, Avijit and Bhatia, Udit},
  journal={npj Urban Sustainability},
  volume={5},
  number={1},
  pages={102},
  year={2025},
  publisher={Nature Publishing Group UK London}
}

@inproceedings{lopez2018microscopic,
  title={Microscopic traffic simulation using sumo},
  author={Lopez, Pablo Alvarez and Behrisch, Michael and Bieker-Walz, Laura and Erdmann, Jakob and Fl{\"o}tter{\"o}d, Yun-Pang and Hilbrich, Robert and L{\"u}cken, Leonhard and Rummel, Johannes and Wagner, Peter and Wie{\ss}ner, Evamarie},
  booktitle={2018 21st international conference on intelligent transportation systems (ITSC)},
  pages={2575--2582},
  year={2018},
  organization={Ieee}
}

@misc{uscensus_tigerline_shapefiles,
  author       = {{U.S. Census Bureau}},
  title        = {TIGER/Line Shapefiles},
  howpublished = {\url{https://www.census.gov/geographies/mapping-files/time-series/geo/tiger-line-file.html}},
  note         = {Accessed 2026-03-17},
  year         = {2026}
}

@misc{openstreetmap_website,
  author       = {{OpenStreetMap contributors}},
  title        = {OpenStreetMap},
  howpublished = {\url{https://www.openstreetmap.org/}},
  note         = {Accessed 2026-03-17},
  year         = {2026}
}

@misc{uscensus_lodes,
  author       = {{U.S. Census Bureau}},
  title        = {LEHD Origin-Destination Employment Statistics (LODES) Data},
  howpublished = {\url{https://lehd.ces.census.gov/data/lodes/}},
  note         = {Accessed 2026-03-17},
  year         = {2026}
}

@misc{acs_b08302,
  author       = {{U.S. Census Bureau}},
  title        = {American Community Survey (ACS) Table B08302: Time Leaving Home to Go to Work},
  howpublished = {\url{https://data.census.gov/table?q=B08302}},
  note         = {ACS 5-year estimates; accessed 2026-03-17},
  year         = {2026}
}

@online{nhts,
  author       = {{National Household Travel Survey}},
  title        = {National Household Travel Survey},
  year         = {2026},
  url          = {https://nhts.ornl.gov/},
  urldate      = {2026-03-17}
}

@article{wing202430,
  title={A 30 m global flood inundation model for any climate scenario},
  author={Wing, Oliver EJ and Bates, Paul D and Quinn, Niall D and Savage, James TS and Uhe, Peter F and Cooper, Anthony and Collings, Thomas P and Addor, Nans and Lord, Natalie S and Hatchard, Simbi and others},
  journal={Water Resources Research},
  volume={60},
  number={8},
  pages={e2023WR036460},
  year={2024},
  publisher={Wiley Online Library}
}

@article{martinez2017new,
  title={A new experiments-based methodology to define the stability threshold for any vehicle exposed to flooding},
  author={Martinez-Gomariz, Eduardo and Gomez, Manuel and Russo, Beniamino and Djordjevi{\'c}, Slobodan},
  journal={Urban Water Journal},
  volume={14},
  number={9},
  pages={930--939},
  year={2017},
  publisher={Taylor \& Francis}
}

@article{pregnolato2017impact,
  title={The impact of flooding on road transport: A depth-disruption function},
  author={Pregnolato, Maria and Ford, Alistair and Wilkinson, Sean M and Dawson, Richard J},
  journal={Transportation research part D: transport and environment},
  volume={55},
  pages={67--81},
  year={2017},
  publisher={Elsevier}
}

@article{pyatkova2019assessing,
  title={Assessing the knock-on effects of flooding on road transportation},
  author={Pyatkova, Katya and Chen, Albert S and Butler, David and Vojinovi{\'c}, Zoran and Djordjevi{\'c}, Slobodan},
  journal={Journal of environmental management},
  volume={244},
  pages={48--60},
  year={2019},
  publisher={Elsevier}
}

@article{girvan2002community,
  title={Community structure in social and biological networks},
  author={Girvan, Michelle and Newman, Mark EJ},
  journal={Proceedings of the national academy of sciences},
  volume={99},
  number={12},
  pages={7821--7826},
  year={2002},
  publisher={The National Academy of Sciences}
}

@article{arrighi2019preparedness,
  title={Preparedness against mobility disruption by floods},
  author={Arrighi, C and Pregnolato, M and Dawson, RJ and Castelli, Fabio},
  journal={Science of the Total Environment},
  volume={654},
  pages={1010--1022},
  year={2019},
  publisher={Elsevier}
}

@article{gangwal2023critical,
  title={Critical facility accessibility and road criticality assessment considering flood-induced partial failure},
  author={Gangwal, Utkarsh and Siders, AR and Horney, Jennifer and Michael, Holly A and Dong, Shangjia},
  journal={Sustainable and Resilient Infrastructure},
  volume={8},
  number={sup1},
  pages={337--355},
  year={2023},
  publisher={Taylor \& Francis}
}

@article{dong2022modest,
  title={Modest flooding can trigger catastrophic road network collapse due to compound failure},
  author={Dong, Shangjia and Gao, Xinyu and Mostafavi, Ali and Gao, Jianxi},
  journal={Communications Earth \& Environment},
  volume={3},
  number={1},
  pages={38},
  year={2022},
  publisher={Nature Publishing Group UK London}
}

@misc{sumoEdgeDataDoc,
  author       = {{German Aerospace Center (DLR)}},
  title        = {SUMO Documentation: Lane- or Edge-based Traffic Measures},
  year         = {2026},
  url          = {https://sumo.dlr.de/docs/Simulation/Output/Lane-_or_Edge-based_Traffic_Measures.html},
  urldate      = {2026-03-18}
}

@article{sanders2023large,
  title={Large and inequitable flood risks in Los Angeles, California},
  author={Sanders, Brett F and Schubert, Jochen E and Kahl, Daniel T and Mach, Katharine J and Brady, David and AghaKouchak, Amir and Forman, Fonna and Matthew, Richard A and Ulibarri, Nicola and Davis, Steven J},
  journal={Nature sustainability},
  volume={6},
  number={1},
  pages={47--57},
  year={2023},
  publisher={Nature Publishing Group UK London}
}

@article{larrabee2022racialized,
  title={Racialized economic segregation and health outcomes: a systematic review of studies that use the Index of Concentration at the Extremes for race, income, and their interaction},
  author={Larrabee Sonderlund, Anders and Charifson, Mia and Schoenthaler, Antoinette and Carson, Traci and Williams, Natasha J},
  journal={PloS one},
  volume={17},
  number={1},
  pages={e0262962},
  year={2022},
  publisher={Public Library of Science San Francisco, CA USA}
}

@incollection{odonnell2008_concentration_index,
  author    = {O'Donnell, Owen and van Doorslaer, Eddy and Wagstaff, Adam and Lindelow, Magnus},
  title     = {The Concentration Index},
  booktitle = {Analyzing Health Equity Using Household Survey Data: A Guide to Techniques and Their Implementation},
  publisher = {The World Bank},
  address   = {Washington, DC},
  year      = {2008},
  chapter   = {8},
  url       = {https://www.worldbank.org/content/dam/Worldbank/document/HDN/Health/HealthEquityCh8.pdf},
  note      = {Accessed 2026-03-18}
}

@article{fajzel2023global,
  title={The global human day},
  author={Fajzel, William and Galbraith, Eric D and Barrington-Leigh, Christopher and Charmes, Jacques and Frie, Elena and Hatton, Ian and Le M{\'e}zo, Priscilla and Milo, Ron and Minor, Kelton and Wan, Xinbei and others},
  journal={Proceedings of the National Academy of Sciences},
  volume={120},
  number={25},
  pages={e2219564120},
  year={2023},
  publisher={National Academy of Sciences}
}

@article{colglazier2015sustainable,
  title={Sustainable development agenda: 2030},
  author={Colglazier, William},
  journal={Science},
  volume={349},
  number={6252},
  pages={1048--1050},
  year={2015},
  publisher={American Association for the Advancement of Science}
}

@article{sun2020dramatic,
  title={Dramatic uneven urbanization of large cities throughout the world in recent decades},
  author={Sun, Liqun and Chen, Ji and Li, Qinglan and Huang, Dian},
  journal={Nature communications},
  volume={11},
  number={1},
  pages={5366},
  year={2020},
  publisher={Nature Publishing Group UK London}
}

@article{zeifman2022world,
  title={A world of 8 billion},
  author={Zeifman, Lubov and Hertog, Sara and Kantorova, Vladimira and Wilmoth, John and others},
  year={2022},
  publisher={United Nations}
}

@article{thacker2019infrastructure,
  title={Infrastructure for sustainable development},
  author={Thacker, Scott and Adshead, Daniel and Fay, Marianne and Hallegatte, St{\'e}phane and Harvey, Mark and Meller, Hendrik and O’Regan, Nicholas and Rozenberg, Julie and Watkins, Graham and Hall, Jim W},
  journal={Nature Sustainability},
  volume={2},
  number={4},
  pages={324--331},
  year={2019},
  publisher={Nature Publishing Group UK London}
}

@article{del2022modeling,
  title={Modeling multimodal access to primary care in an urban environment},
  author={Del Conte, Danielle E and Locascio, Amanda and Amoruso, Joseph and McNamara, Margaret L},
  journal={Transportation Research Interdisciplinary Perspectives},
  volume={13},
  pages={100550},
  year={2022},
  publisher={Elsevier}
}

@article{baeten2018inequalities,
  title={Inequalities in access to healthcare},
  author={Baeten, Rita and Spasova, Slavina and Vanhercke, Bart and Coster, St{\'e}phanie},
  journal={A study of National policies, European social policy network (ESPN). Brussels: European Commission},
  volume={5},
  year={2018}
}

@article{pandey2025rising,
  title={Rising infrastructure inequalities accompany urbanization and economic development},
  author={Pandey, Bhartendu and Brelsford, Christa and Seto, Karen C},
  journal={Nature communications},
  volume={16},
  number={1},
  pages={1193},
  year={2025},
  publisher={Nature Publishing Group UK London}
}

@article{gulliford2002does,
  title={What does' access to health care'mean?},
  author={Gulliford, Martin and Figueroa-Munoz, Jose and Morgan, Myfanwy and Hughes, David and Gibson, Barry and Beech, Roger and Hudson, Meryl},
  journal={Journal of health services research \& policy},
  volume={7},
  number={3},
  pages={186--188},
  year={2002},
  publisher={Sage Publications Sage UK: London, England}
}

@article{levy2010evaluation,
  title={Evaluation of the public health impacts of traffic congestion: a health risk assessment},
  author={Levy, Jonathan I and Buonocore, Jonathan J and Von Stackelberg, Katherine},
  journal={Environmental health},
  volume={9},
  number={1},
  pages={65},
  year={2010},
  publisher={Springer}
}

@article{yiannakoulias2013estimating,
  title={Estimating the effect of turn penalties and traffic congestion on measuring spatial accessibility to primary health care},
  author={Yiannakoulias, Nikolaos and Bland, Widmer and Svenson, Lawrence W},
  journal={Applied Geography},
  volume={39},
  pages={172--182},
  year={2013},
  publisher={Elsevier}
}

@article{lin2024assessing,
  title={Assessing dynamic congestion risks of flood-disrupted transportation network systems through time-variant topological analysis and traffic demand dynamics},
  author={Lin, Xuhui and Lu, Qiuchen and Chen, Long and Brilakis, Ioannis},
  journal={Advanced Engineering Informatics},
  volume={62},
  pages={102672},
  year={2024},
  publisher={Elsevier}
}

@article{tang2018resilience,
  title={A resilience-oriented approach for quantitatively assessing recurrent spatial-temporal congestion on urban roads},
  author={Tang, Junqing and Heinimann, Hans Rudolf},
  journal={PloS one},
  volume={13},
  number={1},
  pages={e0190616},
  year={2018},
  publisher={Public Library of Science San Francisco, CA USA}
}

@article{schuster2024stress,
  title={Stress-testing road networks and access to medical care},
  author={Schuster, Hannah and Polleres, Axel and Wachs, Johannes},
  journal={Transportation research part A: policy and practice},
  volume={181},
  pages={104017},
  year={2024},
  publisher={Elsevier}
}

@article{gangwal2022critical,
  title={Critical facility accessibility rapid failure early-warning detection and redundancy mapping in urban flooding},
  author={Gangwal, Utkarsh and Dong, Shangjia},
  journal={Reliability Engineering \& System Safety},
  volume={224},
  pages={108555},
  year={2022},
  publisher={Elsevier}
}

@article{alabbad2021assessment,
  title={Assessment of transportation system disruption and accessibility to critical amenities during flooding: Iowa case study},
  author={Alabbad, Yazeed and Mount, Jerry and Campbell, Ann Melissa and Demir, Ibrahim},
  journal={Science of the total environment},
  volume={793},
  pages={148476},
  year={2021},
  publisher={Elsevier}
}

@article{pescaroli2018perceptions,
  title={Perceptions of cascading risk and interconnected failures in emergency planning: Implications for operational resilience and policy making},
  author={Pescaroli, Gianluca},
  journal={International journal of disaster risk reduction},
  volume={30},
  pages={269--280},
  year={2018},
  publisher={Elsevier}
}

@article{serre2018assessing,
  title={Assessing and mapping urban resilience to floods with respect to cascading effects through critical infrastructure networks},
  author={Serre, Damien and Heinzlef, Charlotte},
  journal={International Journal of Disaster Risk Reduction},
  volume={30},
  pages={235--243},
  year={2018},
  publisher={Elsevier}
}

@article{liu2023beyond,
  title={Beyond residence: A mobility-based approach for improved evaluation of human exposure to environmental hazards},
  author={Liu, Zhewei and Liu, Chenyue and Mostafavi, Ali},
  journal={Environmental Science \& Technology},
  volume={57},
  number={41},
  pages={15511--15522},
  year={2023},
  publisher={ACS Publications}
}

@article{huang2025human,
  title={Human mobility under disasters: a systematic review and framework for equitable and resilient mobility governance},
  author={Huang, Fengjue and Tang, Junqing and Zhao, Pengjun and Chen, Zhihe and Li, Jiaying and Lyu, Wei},
  journal={npj Natural Hazards},
  volume={2},
  number={1},
  pages={99},
  year={2025},
  publisher={Nature Publishing Group UK London}
}

@article{yu2020disruption,
  title={Disruption of emergency response to vulnerable populations during floods},
  author={Yu, Dapeng and Yin, Jie and Wilby, Robert L and Lane, Stuart N and Aerts, Jeroen CJH and Lin, Ning and Liu, Min and Yuan, Hongyong and Chen, Jianguo and Prudhomme, Christel and others},
  journal={Nature Sustainability},
  volume={3},
  number={9},
  pages={728--736},
  year={2020},
  publisher={Nature Publishing Group UK London}
}

@article{rajput2023anatomy,
  title={Anatomy of perturbed traffic networks during urban flooding},
  author={Rajput, Akhil Anil and Nayak, Sanjay and Dong, Shangjia and Mostafavi, Ali},
  journal={Sustainable Cities and Society},
  volume={97},
  pages={104693},
  year={2023},
  publisher={Elsevier}
}

@article{garcia2025future,
  title={Future-proofing cities against negative city mobility and public health impacts of impending natural hazards: a system dynamics modelling study},
  author={Garcia, Leandro and Hafezi, Mehdi and Lima, Larissa and Millett, Christopher and Thompson, Jason and Wang, Ruoyu and Akaraci, Selin and Goel, Rahul and Reis, Rodrigo and Nice, Kerry A and others},
  journal={The Lancet Planetary Health},
  volume={9},
  number={3},
  pages={e207--e218},
  year={2025},
  publisher={Elsevier}
}

@misc{fathom_us_flood_risk_index_misc,
  author       = {{Fathom}},
  title        = {US Flood Risk Index},
  howpublished = {\url{https://www.fathom.global/us-flood-risk-index/}},
  note         = {Accessed 2026-04-09},
  year         = {n.d.}
}

@article{koks2019global,
  title={A global multi-hazard risk analysis of road and railway infrastructure assets},
  author={Koks, Elco E and Rozenberg, Julie and Zorn, Conrad and Tariverdi, Mersedeh and Vousdoukas, Michalis and Fraser, Stuart Alexander and Hall, JW and Hallegatte, Stephane},
  journal={Nature communications},
  volume={10},
  number={1},
  pages={2677},
  year={2019},
  publisher={Nature Publishing Group UK London}
}

@article{papilloud2024road,
  title={Road network disruptions during extreme flooding events and their impact on the access to emergency medical services: A spatiotemporal vulnerability analysis},
  author={Papilloud, Tsolmongerel and Steiner, Albert and Zischg, Andreas and Keiler, Margreth},
  journal={Science of the Total Environment},
  volume={956},
  pages={177140},
  year={2024},
  publisher={Elsevier}
}

@article{li2025measuring,
  title={Measuring emergency medical service accessibility using the improved 3SFCA: With a focus on key population influence},
  author={Li, Chuanyao and Jia, Baibing},
  journal={Journal of Transport Geography},
  volume={128},
  pages={104310},
  year={2025},
  publisher={Elsevier}
}

@article{zeng2025unveiling,
  title={Unveiling city jam-prints of urban traffic based on jam patterns},
  author={Zeng, Guanwen and Serok, Nimrod and Lieberthal, Efrat Blumenfeld and Duan, Jinxiao and Liu, Shiyan and Sui, Shaobo and Li, Daqing and Havlin, Shlomo},
  journal={Communications Physics},
  volume={8},
  number={1},
  pages={121},
  year={2025},
  publisher={Nature Publishing Group UK London}
}

@report{ASCE_IRC_2025,
  author       = {{American Society of Civil Engineers}},
  title        = {2025 Report Card for America's Infrastructure: Full Report},
  year         = {2025},
  month        = mar,
  institution  = {American Society of Civil Engineers},
  url          = {https://infrastructurereportcard.org/wp-content/uploads/2025/03/Full-Report-2025-Natl-IRC-WEB.pdf},
  }

@article{he2021flood,
  title={Flood impacts on urban transit and accessibility—A case study of Kinshasa},
  author={He, Yiyi and Thies, Stephan and Avner, Paolo and Rentschler, Jun},
  journal={Transportation research part D: transport and environment},
  volume={96},
  pages={102889},
  year={2021},
  publisher={Elsevier}
}

@article{maymandi2022compound,
  title={Compound coastal, fluvial, and pluvial flooding during historical hurricane events in the Sabine--Neches Estuary, Texas},
  author={Maymandi, Nahal and Hummel, Michelle A and Zhang, Yu},
  journal={Water Resources Research},
  volume={58},
  number={12},
  pages={e2022WR033144},
  year={2022},
  publisher={Wiley Online Library}
}

@article{wahl2015increasing,
  title={Increasing risk of compound flooding from storm surge and rainfall for major US cities},
  author={Wahl, Thomas and Jain, Shaleen and Bender, Jens and Meyers, Steven D and Luther, Mark E},
  journal={Nature Climate Change},
  volume={5},
  number={12},
  pages={1093--1097},
  year={2015},
  publisher={Nature Publishing Group UK London}
}

@article{hendry2019assessing,
  title={Assessing the characteristics and drivers of compound flooding events around the UK coast},
  author={Hendry, Alistair and Haigh, Ivan D and Nicholls, Robert J and Winter, Hugo and Neal, Robert and Wahl, Thomas and Joly-Laugel, Am{\'e}lie and Darby, Stephen E},
  journal={Hydrology and Earth System Sciences},
  volume={23},
  number={7},
  pages={3117--3139},
  year={2019},
  publisher={Copernicus Publications G{\"o}ttingen, Germany}
}

@article{kasmalkar2020floods,
  title={When floods hit the road: Resilience to flood-related traffic disruption in the San Francisco Bay Area and beyond},
  author={Kasmalkar, Indraneel G and Serafin, Katherine A and Miao, Yufei and Bick, I Avery and Ortolano, Leonard and Ouyang, Derek and Suckale, Jenny},
  journal={Science advances},
  volume={6},
  number={32},
  pages={eaba2423},
  year={2020},
  publisher={American Association for the Advancement of Science}
}

@article{sohrabi2025analyzing,
  title={Analyzing compound flooding drivers across the US Gulf Coast States},
  author={Sohrabi, Meraj and Moftakhari, Hamed and Moradkhani, Hamid},
  journal={Geophysical Research Letters},
  volume={52},
  number={9},
  pages={e2025GL114769},
  year={2025},
  publisher={Wiley Online Library}
}

@article{ali2025multivariate,
  title={Multivariate compound events drive historical floods and associated losses along the US East and Gulf coasts},
  author={Ali, Javed and Wahl, Thomas and Morim, Joao and Enriquez, Alejandra and Gall, Melanie and Emrich, Christopher T},
  journal={npj Natural Hazards},
  volume={2},
  number={1},
  pages={19},
  year={2025},
  publisher={Nature Publishing Group UK London}
}

@article{frei2006future,
  title={Future change of precipitation extremes in Europe: Intercomparison of scenarios from regional climate models},
  author={Frei, Christoph and Sch{\"o}ll, Regina and Fukutome, Sophie and Schmidli, J{\"u}rg and Vidale, Pier Luigi},
  journal={Journal of Geophysical Research: Atmospheres},
  volume={111},
  number={D6},
  year={2006},
  publisher={Wiley Online Library}
}

@article{korkali2017reducing,
  title={Reducing cascading failure risk by increasing infrastructure network interdependence},
  author={Korkali, Mert and Veneman, Jason G and Tivnan, Brian F and Bagrow, James P and Hines, Paul DH},
  journal={Scientific reports},
  volume={7},
  number={1},
  pages={44499},
  year={2017},
  publisher={Nature Publishing Group UK London}
}

@article{buchanan2017amplification,
  title={Amplification of flood frequencies with local sea level rise and emerging flood regimes},
  author={Buchanan, Maya K and Oppenheimer, Michael and Kopp, Robert E},
  journal={Environmental Research Letters},
  volume={12},
  number={6},
  pages={064009},
  year={2017},
  publisher={IOP Publishing}
}

@article{slater2016recent,
  title={Recent trends in US flood risk},
  author={Slater, Louise J and Villarini, Gabriele},
  journal={Geophysical Research Letters},
  volume={43},
  number={24},
  pages={12--428},
  year={2016},
  publisher={Wiley Online Library}
}

@article{weiss2018global,
  title={A global map of travel time to cities to assess inequalities in accessibility in 2015},
  author={Weiss, D J and Nelson, Andy and Gibson, HS and Temperley, W and Peedell, Stephen and Lieber, Allie and Hancher, Matt and Poyart, Eduardo and Belchior, Sim{\~a}o and Fullman, Nancy and others},
  journal={Nature},
  volume={553},
  number={7688},
  pages={333--336},
  year={2018},
  publisher={Nature Publishing Group UK London}
}

@article{meijer2018global,
  title={Global patterns of current and future road infrastructure},
  author={Meijer, Johan R and Huijbregts, Mark AJ and Schotten, Kees CGJ and Schipper, Aafke M},
  journal={Environmental Research Letters},
  volume={13},
  number={6},
  pages={064006},
  year={2018},
  publisher={IOP Publishing}
}

@article{abbiasov202415,
  title={The 15-minute city quantified using human mobility data},
  author={Abbiasov, Timur and Heine, Cate and Sabouri, Sadegh and Salazar-Miranda, Arianna and Santi, Paolo and Glaeser, Edward and Ratti, Carlo},
  journal={Nature Human Behaviour},
  volume={8},
  number={3},
  pages={445--455},
  year={2024},
  publisher={Nature Publishing Group UK London}
}

@article{marzolla2026proximity,
  title={Proximity-based cities emit less mobility-driven CO2},
  author={Marzolla, Francesco and M. Melo, Hygor P and Bruno, Matteo and Loreto, Vittorio},
  journal={npj Sustainable Mobility and Transport},
  volume={3},
  number={1},
  pages={7},
  year={2026},
  publisher={Nature Publishing Group UK London}
}

@article{ma202315,
  title={15-min pedestrian distance life circle and sustainable community governance in Chinese metropolitan cities: A diagnosis},
  author={Ma, Wenjun and Wang, Ning and Li, Yuxi and Sun, Daniel Jian},
  journal={Humanities and Social Sciences Communications},
  volume={10},
  number={1},
  pages={364},
  year={2023},
  publisher={Palgrave}
}

@article{lee2025travel,
  title={Travel efficiency in urban space: the role of built environment in shaping excess travel distance across transport modes},
  author={Lee, Sangwan and Cho, Kuk and Jeon, Yonghyen},
  journal={Scientific Reports},
  volume={15},
  number={1},
  pages={33372},
  year={2025},
  publisher={Nature Publishing Group UK London}
}

@article{zhao2025unequal,
  title={Unequal roles of cities in the intercity healthcare system},
  author={Zhao, Pengjun and Li, Juan and Zhang, Mengzhu},
  journal={Nature Cities},
  volume={2},
  number={3},
  pages={198--209},
  year={2025},
  publisher={Nature Publishing Group US New York}
}

@article{wang2019local,
  title={Local floods induce large-scale abrupt failures of road networks},
  author={Wang, Weiping and Yang, Saini and Stanley, H Eugene and Gao, Jianxi},
  journal={Nature communications},
  volume={10},
  number={1},
  pages={2114},
  year={2019},
  publisher={Nature Publishing Group UK London}
}

@article{dave2021extreme,
  title={Extreme precipitation induced concurrent events trigger prolonged disruptions in regional road networks},
  author={Dave, Raviraj and Subramanian, Srikrishnan Siva and Bhatia, Udit},
  journal={Environmental Research Letters},
  volume={16},
  number={10},
  pages={104050},
  year={2021},
  publisher={IOP Publishing}
}

@article{hines2023hospital,
  title={Hospital preparedness, mitigation, and response to Hurricane Harvey in Harris County, Texas},
  author={Hines, Emmanuelle and Reid, Colleen E},
  journal={Disaster medicine and public health preparedness},
  volume={17},
  pages={e18},
  year={2023},
  publisher={Cambridge University Press}
}

@article{uppal2013search,
  title={In search of the silver lining. The impact of Superstorm Sandy on Bellevue Hospital},
  author={Uppal, Amit and Evans, Laura and Chitkara, Nishay and Patrawalla, Paru and Mooney, M Ann and Addrizzo-Harris, Doreen and Leibert, Eric and Reibman, Joan and Rogers, Linda and Berger, Kenneth I and others},
  journal={Annals of the American Thoracic Society},
  volume={10},
  number={2},
  pages={135--142},
  year={2013},
  publisher={Oxford University Press}
}

@article{xu2024unified,
  title={A unified dataset for the city-scale traffic assignment model in 20 US cities},
  author={Xu, Xiaotong and Zheng, Zhenjie and Hu, Zijian and Feng, Kairui and Ma, Wei},
  journal={Scientific data},
  volume={11},
  number={1},
  pages={325},
  year={2024},
  publisher={Nature Publishing Group UK London}
}

@article{smith2020coastal,
  title={Coastal amenities and income stratification},
  author={Smith, V Kerry and Whitmore, Ben},
  journal={Economics Letters},
  volume={192},
  pages={109241},
  year={2020},
  publisher={Elsevier}
}

@book{un2018urbanization,
  author    = {{United Nations, Department of Economic and Social Affairs, Population Division}},
  title     = {World Urbanization Prospects: The 2018 Revision},
  year      = {2018},
  address    = {New York},
  publisher = {United Nations}
}

@article{salas2024impact,
  title={Impact of extreme weather events on healthcare utilization and mortality in the United States},
  author={Salas, Renee N and Burke, Laura G and Phelan, Jessica and Wellenius, Gregory A and Orav, E John and Jha, Ashish K},
  journal={Nature Medicine},
  volume={30},
  number={4},
  pages={1118--1126},
  year={2024},
  publisher={Nature Publishing Group US New York}
}

@article{zhong2025universal,
  title={Universal expansion of human mobility across urban scales},
  author={Zhong, Lu and Dong, Lei and Wang, Qi R and Song, Chaoming and Gao, Jianxi},
  journal={Nature Cities},
  volume={2},
  number={7},
  pages={603--607},
  year={2025},
  publisher={Nature Publishing Group US New York}
}

@misc{census_commuting_acs_2025,
  author       = {{U.S. Census Bureau}},
  title        = {United States Commuting At A Glance: American Community Survey 1-Year Estimates},
  year         = {2025},
  month        = sep,
  howpublished = {\url{https://www.census.gov/topics/employment/commuting/guidance/acs-1yr.html}},
  note         = {Page last revised September 22, 2025; accessed 2026-04-14}
}

@article{martello2021evaluation,
  title={Evaluation of climate change resilience for Boston’s rail rapid transit network},
  author={Martello, Michael V and Whittle, Andrew J and Keenan, Jesse M and Salvucci, Frederick P},
  journal={Transportation Research Part D: Transport and Environment},
  volume={97},
  pages={102908},
  year={2021},
  publisher={Elsevier}
}

@article{del2025composite,
  title={A composite index framework for compound flood risk assessment},
  author={Del-Rosal-Salido, Juan and Berm{\'u}dez, Mar{\'\i}a and Ortega-S{\'a}nchez, Miguel and Sanuy, Marc and Silva-Santana, Marcus and Jim{\'e}nez, Jos{\'e} A},
  journal={Communications Earth \& Environment},
  volume={6},
  number={1},
  pages={342},
  year={2025},
  publisher={Nature Publishing Group UK London}
}

@article{flanagan2011social,
  title={A social vulnerability index for disaster management},
  author={Flanagan, Barry E and Gregory, Edward W and Hallisey, Elaine J and Heitgerd, Janet L and Lewis, Brian},
  journal={Journal of Homeland Security and Emergency Management},
  volume={8},
  year={2011}
}






\end{document}